\documentclass[%
 reprint,
showpacs,
nofootinbib,
nobibnotes,
 amsmath,amssymb,
 aps,
]{revtex4-1}
\usepackage{graphicx}
\usepackage{bm}
\usepackage{hyperref}
\usepackage[mathlines]{lineno}

\begin{document}
\preprint{APS/123-QED}
\title{Evolution of magnetic fields through cosmological perturbation theory}
\author{H\'ector J. Hortua}
 \email{hjhortuao@unal.edu.co}
\author{Leonardo Casta\~neda}%
 \email{lcastanedac@unal.edu.co}
\author{J. M. Tejeiro}%
 \email{jmtejeiros@unal.edu.co}
\affiliation{%
 Grupo de Gravitaci\'on y Cosmolog\'ia, Observatorio Astron\'omico Nacional,\\
 Universidad Nacional de Colombia, cra 45 $\#$26-85, Ed. Uriel Gutierr\'ez, Bogot\'a D.C, Colombia
}%



\date{\today}

\begin{abstract}
The origin of galactic and extra-galactic magnetic fields is an unsolved problem in  modern cosmology. A possible scenario comes from the idea of these fields emerged from a small field, \textit{a seed}, which was produced in the early universe (phase transitions, inflation, ...) and it evolves in time. Cosmological perturbation theory offers a natural way to study the evolution of primordial magnetic fields. The dynamics for this field in the cosmological context is described by a cosmic dynamo like equation, through the dynamo term. In this paper we get the  perturbed Maxwell's equations  and compute the energy momentum tensor to second order in perturbation theory in terms of gauge invariant quantities. Two posible scenarios are discussed, first we consider a FLRW background without magnetic field and we study the perturbation theory introducing the magnetic field as a perturbation. The second scenario, we consider a magnetized FLRW and build up  the perturbation theory from this background. We compare the cosmological dynamo like equation in both scenarios. 
\begin{description}
\item[PACS numbers]
98.80.-k, 95.30.Qd.
\end{description}
\end{abstract}
\pacs{Valid PACS appear here}
\maketitle


\section{Introduction}\label{sec:level1}
Magnetic fields  have been observed on several scales in the universe. Galaxies and clusters of galaxies contain magnetic fields with strengths of $\sim 10^{-6}$ G  \cite{1a,*1b}, fields within clusters are also likely to exist, with  strengths of  comparable magnitude \cite{gio1,*1}. There is also evidence of magnetic fields on scales of superclusters \cite{feyn54,*3,*4,*5}. On the other hand, the possibility of  cosmological  magnetic field has been addressed  comparing the CMB quadrupole with  one induced by a constant magnetic field (in cohe\-rence scales of $\sim1Mpc$), constraining the field magnitude to $ B<6.8\times 10^{-9} (\Omega_{m}h^{2})^{1/2}$ Gauss \cite{6,*7,*8}.
Ho\-wever, the origin of such large scale magnetic fields  is still unknown. These fields are assumed to be increased and maintained by dynamo mechanism, but  it needs a \textit{seed} before the mecha\-nism takes place \cite{8a}. Astrophysical mecha\-nisms, such  as the Biermann battery   have been used to explain how the magnetic field is mantained  in  objects such  as gala\-xies, stars and supernova remnants \cite{9,*10,*11,*12}, but they are not likely  correlated  beyond galactic sizes \cite{13}. It makes difficult to use astrophysical mechanisms to explain the origin of magnetic fields on cosmological scales. In order to overcome this problem, the primordial origin should be found in other scenarios from  which the astrophysical mechanism starts.   For example, magnetic fields could be ge\-nerated during primordial phase transitions (such as QCD, the electroweak or GUT), parity-violating processes that generates magnetic helicity or during inflation \cite{14,*15,*16,*17,*18,*18a,*18b,*19,*20,*201}. Magnetic fields also are gene\-rated during the radiation era in regions with nonvanishing vorticity.  This seed  was proposed by Harrison \cite{20a,*20b,*20bc}. Magnetic fields genera\-tion from density fluctuations in pre-recombination era has been investigated in \cite{20c}.  The advantage of these primordial processes is that they offer a wide range of coherence lengths (many of which are strongly constrained by Nucleosynthesis \cite{21,*Tinaa1,*Tinab1}), while the astrophysical mechanisms  produce fields at the same order of the   astrophysical size of the  object. Recently  a lower limit of the large scale correlated magnetic field was found. It constrains models for the origin of cosmic magnetic fields, giving a possible evidence for their primordial origin \cite{Tina11,*Tina12,*Tina13}.\\     
One way to describe the evolution of magnetic fields is through Cosmological Perturbation Theory. This theo\-ry \cite{22,*22a,*22b} is a  powerful  tool for  understanding the present properties of the large-scale structure of the Universe and their origin. It has been mainly used to predict e\-ffects on the temperature distribution in the  Cosmic Microwave Background (CMB) \cite{23,*24}. Futhermore, li\-near perturbation theory combined with  inflation suggests  that  primordial fluctuations of the universe are adiabatic and Gaussian \cite{25}. However, due to the high precision measurements reached  in  cosmology, higher order  cosmological perturbation theory is required to test the cu\-rrent cosmological framework \cite{26}\cite{29new1}.
There are mainly two approaches to studying higher order perturbative e\-ffects: one uses nonlinear theory and different manifestations of the separate universe approximation, using the $\Delta N$ formalism \cite{27,28},  and the other  is the  Bardeen  approach where metric and  matter fields are expanded in a power series \cite{bar}. Within the Bardeen approach, a set of variables are determined in such a way that has no gauge dependence. These are known in the literature as \textit{gauge-invariant variables} which have been widely used in different cosmological  scenarios \cite{29new2}. One important result of cosmological perturbation theory is the coupling between gravity and electromagnetic fields,  which have shown a  magneto-geometrical interaction that could change the evolution of the fields on large scales. An effect\- is the  amplification of cosmic fields. Indeed,  large scale magnetic fields in perturbed spatially open FLRW models decay as $a^{-1}$, a rate considerably slower than the standard  $a^{-2}$ \cite{29b,*29a,*29c,*29d}.  The hyperbolic geometry of these open FLRW models leads to the superadiabatic amplification on large scales \cite{29e}.  
The main goal in this paper is to study the late evolution  of magnetic fields  that were generated in early stages of the universe. We use the cosmological perturbation theory following the Gauge Invariant  formalism to find  the perturbed Maxwell equations up to second order and  also we obtain a dynamo like equation written in terms of gauge invariant variables to first and second order. Futhermore,  we discuss the importance that both curvature and the gravi\-tational potential plays in the evolution of these fields. The paper is organized as follows: in the next section we briefly give an introduction of cosmological perturbation theory and  we address the gauge problem  in this theo\-ry.
 The section \ref{background}  presents the matter equations in the  homogeneous and isotropic universe, which was used to generate  the first and second order dynamical equations. In section \ref{GI}, we define the first  order gauge invariant variables for the  perturbations not only in the matter (energy density, presure, magnetic and electric field) but also in the geo\-metrical quantities (gravitational potential, curvature, shear ..).  The first-order perturbation of the Maxwell's equations is reviewed in section \ref{2sec} and together with  the  Ohm's law allows  to find the cosmological dynamo equation to describe the evolution of the magnetic field.  The derivation of  second-order Maxwell's equations is given in section \ref{sec:3}, and following the same methodology for the first-order case, we find the cosmological dynamo equation at second order written in terms of gauge invariant variables. In the section \ref{2part}, we use  an alternative approximation to the model considering a magnetic field in the FLRW background. It is found that  amplification effects of magnetic field appear at first order in the equations, besides of the absence of fractional orders. Also a discussion  between both approaches is done. The final section \ref{conclusi}   is devoted to a discussion of  the main results  and the connection with future works.

\section{The gauge problem in perturbation theory}
Perturbation 
theory  helps us
to  find  approximate solutions of the 
Einstein field equations through small desviations from  an exact solution 
\cite{30}.
 In this theory  one works with two differents space-times, one is the real space-time  $(\mathcal{M},g_{\alpha \beta})$  which  describes the perturbed universe and the other is the background space-time $(\mathcal{M}_{0},g_{\alpha \beta}^{(0)})$  which is an idealization  and  is taken as reference  to generate the real space-time. Then, the perturbation of any  quantity $\varGamma$ (e.g., energy density $\mu(x,t)$,  4-velocity $u^{\alpha}(x,t)$, magnetic field $B^{i}(x,t)$ or metric tensor ${g}_{\alpha \beta}$) is the difference between the value  that the quantity $\varGamma$ takes  in the real space-time  and the value in the background at a given point \footnote{This difference should be taken in the same physical point.}.
In order  to determinate the perturbation in $\varGamma$, we must have a way to compare $\varGamma$ (tensor on the real space-time) with $\varGamma^{(0)}$ (being $\varGamma^{(0)}$ the value on $\mathcal{M}_{0}$). This requires the assumption to identify points of $\mathcal{M}$ with those of $\mathcal{M}_{0}$. This is  accomplished by assigning a mapping between these space-times  called \textit{gauge choice} given by a function   $\mathcal{X}:\:\mathcal{M}_{0}(p)\longrightarrow\mathcal{M}(\bar{p})$ for any  point $p\in \mathcal{M}_{0}$ and $\bar{p}\in \mathcal{M}$, which generate a pull-back
\begin{equation}
\begin{array}{c}
\mathcal{X}^{*}:\\
\\\end{array}\;\begin{array}{c}
\mathcal{M}\\
T^{*}(\overline{p})\end{array}\longrightarrow\begin{array}{c}
\mathcal{M}_{0}\\
T^{*}(p)\end{array},\end{equation}
thus, points on the real and background space-time can be compared through of $\mathcal{X}$. Then, the  perturbation for $\varGamma$ is defined as
\begin{equation}
  \delta\varGamma(p)=\varGamma(\bar{p})-\varGamma^{(0)}(p)\label{perturbation}.
\end{equation}
We see that the perturbation $\delta\varGamma$ is completely dependent of the gauge choice because the mapping determines the representation on $\mathcal{M}_{0}$ of  $\varGamma(\bar{p})$.   However, one can also choose another correspondence $\mathcal{Y}$ between these space-times so that $\mathcal{Y}:\,\mathcal{M}_{0}(q)\rightarrow\mathcal{M}(\overline{p})$, ($p\neq q$).\footnote{This is the active approach where  transformations of the perturbed quantities are evaluated at the same coordinate point.} In the literature a change of this identification map  is called \textit{gauge transformation}.  The  freedom  to choose between different correspondences is due to the general covariance in General Relativity, which states that there is no preferred coordinate system in nature \cite{sach,*31}. 
Hence, this freedom   will generate an  arbitrariness in the value of $\delta\varGamma$ at any space-time point $p$, which is called  \textit{gauge problem} in the general relativistic perturbation theory and has been treated by  \cite{ellis,*nak}.   This problem generates unphysical degree of freedom to the solutions in the theory and  therefore one  should  fix the gauge  or  build up nondependent quantities of the gauge.

\subsection{Gauge transformations and gauge invariant variables}
 To define the perturbation to a given order, it is ne\-cessary to introduce the concept of Taylor expansion on a manifold and thus  the metric and  matter fields are expanded in a power series. Following \cite{stewar,*mollerach,*nakamuranuevo},  is considered a family of four-dimensional submanifolds $\mathcal{M}_{\lambda}$ with $\lambda\in\mathbb{R}$, embedded in a 5-dimensional manifold $\mathcal{N}=\mathcal{M}\times\mathbb{R}$. Each submanifold in the family represents a perturbed space-time and the background space-time  is represented by the manifold $\mathcal{M}_{0}$ ($\lambda=0$). On these mani\-folds we consider that the Einstein field and Maxwell's equations are satisfied
\begin{equation}
\texttt{E}[g_{\lambda},T_{\lambda}]=0\quad \text{and}\quad{M}[F_{\lambda},J_{\lambda}]=0; \label{suposicion maxwellyeinsteinmanifol}
\end{equation}
each tensor field $\varGamma_{\lambda}$ on a given manifold $\mathcal{M}_{\lambda}$ is extended to all manifold $\mathcal{N}$ through  $\varGamma(p,\lambda)\equiv \varGamma_{\lambda}(p)$ to any $p\in\mathcal{M}_{\lambda}$ likewise the above equations are extended to $\mathcal{N}$.\footnote{In eq. (\ref{suposicion maxwellyeinsteinmanifol}),  $g_{\lambda}$ and $T_{\lambda}$ are the metric and the matter fields on $\mathcal{M}_{\lambda}$, similarly $F_{\lambda}$ and $J_{\lambda}$ are the electromagnetic field and the four-current on $\mathcal{M}_{\lambda}$.} 
We  used a diffeomorphism such that the difference in the right side of eq. (\ref{perturbation}) can be done. Is introduced an one-parameter group of diffeomorphisms $\mathcal{X}_{\lambda}$ which identifies points in the background with points in the real space-time labeled with the value $\lambda$.  Each $\mathcal{X}_{\lambda}$ is  a member of a flow $\mathcal{X}$ on $\mathcal{N}$  and it specifies a vector field $X$ with the property $X^{4}=1$ everywhere (transverse to the $\mathcal{M}_{\lambda}$)\footnote{Here we introduce a coordinate system $x^{\alpha}$ through a chart on $\mathcal{M}_{\lambda}$ with $\alpha=0,1,2,3$, thus, giving a vector field on $\mathcal{N}$, which has the property that $X^{4}=\lambda$ in this chart, while the other components remain arbitrary.} then points which lie on the same integral curve  of $X$ have to be regarded as the same point \cite{nak}. Therefore, according to the above, one gets a definition for the tensor perturbation
\begin{equation}
\Delta\varGamma_{\lambda}\equiv \mathcal{X}_{\lambda}^{*}\varGamma |_{\mathcal{M}_{0}}-\varGamma_{0}.\label{eq:20}\end{equation}
At higher orders the Taylor expansion is given by \cite{mollerach},
\begin{equation}
\Delta^{\mathcal{X}}\varGamma_{\lambda}=\sum_{k=0}^{\infty}\frac{\lambda^{k}}{k!}\delta^{(k)}_{\mathcal{X}}\varGamma-\varGamma_{0}=\sum_{k=1}^{\infty}\frac{\lambda^{k}}{k!}\delta^{(k)}_{\mathcal{X}}\varGamma,
\label{3.3}\end{equation}
where
\begin{equation}
\delta^{(k)}_{\mathcal{X}}\varGamma=\left[\frac{d^{k}\mathcal{X}_{\lambda}^{*}\varGamma}{d\lambda^{k}}\right]_{\lambda=0,\mathcal{M}_{0}}.\label{eq:21}
\end{equation}
Now, rewriting eq. (\ref{eq:20}) we get
\begin{equation}
\mathcal{X}_{\lambda}^{*}\varGamma|_{\mathcal{M}_{0}}=\varGamma_{0}+ \lambda\delta^{(1)}_{\mathcal{X}}\varGamma+\frac{\lambda^{2}}{2}\delta^{(2)}_{\mathcal{X}}\varGamma+\mathcal{O}(\lambda^{3}),
 \label{split}\end{equation}
Notice in the eqs. (\ref{eq:21}) and (\ref{split}) the representation of $\varGamma$ on $\mathcal{M}_{0}$ is splitting in the background value  $\varGamma_{0}$ plus $\mathcal{O}(k)$ perturbations  in the gauge $\mathcal{X}_{\lambda}$. Therefore,   the  $k$-th order $\mathcal{O}(k)$ in  $\varGamma$ depends on gauge ${\mathcal{X}}$. With this description the “perturbations are fields lie in the background”. The first term  in  eq. (\ref{eq:20}) admits an expansion around  $\lambda=0$ given by \cite{mollerach} 
\begin{equation}
\mathcal{X}_{\lambda}^{*}\varGamma |_{\mathcal{M}_{0}}=\sum_{k=0}^{\infty}\frac{\lambda^{k}}{k!}\mathcal{L}^{k}_{X}\varGamma|_{\mathcal{M}_{0}}=\exp{(\lambda\mathcal{L}_{X})}\varGamma|_{\mathcal{M}_{0}},
\label{23}\end{equation}
where $\mathcal{L}_{X}\varGamma$ is the Lie derivative of $\varGamma$ with respect to a vector field $X$ that  generates the flow $\mathcal{X}$.  If we define $\mathcal{X}_{\lambda}^{*}\varGamma|_{\mathcal{M}_{0}}\equiv\varGamma_{\lambda}^{\mathcal{X}}$ and proceeding in the same way  for another gauge choice $\mathcal{Y}$, using eqs. (\ref{eq:20})-(\ref{23}), the tensor fields $\varGamma_{\lambda}^{\mathcal{X,Y}}$ can be written as
\begin{equation}
\varGamma_{\lambda}^{\mathcal{X}}=\sum_{k=0}^{\infty}\frac{\lambda^{k}}{k!}\delta^{(k)}_{\mathcal{X}}\varGamma=\sum_{k=0}^{\infty}\frac{\lambda^{k}}{k!}\mathcal{L}^{k}_{X}\varGamma|_{\mathcal{M}_{0}},\label{a}
\end{equation}
\begin{equation}
 \varGamma_{\lambda}^{\mathcal{Y}}=\sum_{k=0}^{\infty}\frac{\lambda^{k}}{k!}\delta^{(k)}_{\mathcal{Y}}\varGamma=\sum_{k=0}^{\infty}\frac{\lambda^{k}}{k!}\mathcal{L}^{k}_{Y}\varGamma|_{\mathcal{M}_{0}},\label{b}
\end{equation}
if $\varGamma_{\lambda}^{\mathcal{X}}=\varGamma_{\lambda}^{\mathcal{Y}}$ for any  arbitrary gauge ${\mathcal{X}}$ and ${\mathcal{Y}}$, from here it is clear that  $\varGamma$ is totally gauge invariant.
It is also clear  that $\varGamma$ is gauge invariant  to order $n\geqslant1$   if only if satisfy $\delta^{(k)}_{\mathcal{Y}}\varGamma=\delta^{(k)}_{\mathcal{X}}\varGamma$, or  in other way
\begin{equation}
 \mathcal{L}_{X}\delta^{(k)}\varGamma=0,\label{IG}
\end{equation}
for any vector field ${X}$ and $\forall k<n$. To first order ($k=1$) any scalar that is  constant in the background or any tensor that vanished in the background are gauge invariant. This result is  known  as  \textit{Stewart-Walker Lemma} \cite{lemma}, i.e.,  eq. (\ref{IG}) generalizes  this Lemma.
However, when $\varGamma$ is not gauge invariant and  there are two   gauge choices  $\mathcal{X}_{\lambda}$,  $\mathcal{Y}_{\lambda}$, the representation of  $\varGamma|_{\mathcal{M}_{0}}$ is different depending of the used gauge. To transform the representation from a gauge choice $\mathcal{X}_{\lambda}^{*}\varGamma|_{\mathcal{M}_{0}}$ to another $\mathcal{Y}_{\lambda}^{*}\varGamma|_{\mathcal{M}_{0}}$ as  with  the map  $\Phi_{\lambda}:{\mathcal{M}_{0}}\rightarrow{\mathcal{M}_{0}}$ given by
\begin{equation}
 \Phi_{\lambda}\equiv\mathcal{X}_{-\lambda}\circ\mathcal{Y}_{\lambda}\Rightarrow\varGamma_{\lambda}^{\mathcal{Y}}=\Phi_{\lambda}^{*}\varGamma_{\lambda}^{\mathcal{X}},\label{phi}
\end{equation}
as a consecuence, the diffeomorphism $\Phi_{\lambda}$ induce a pull-back $\Phi_{\lambda}^{*}$ which changes the representation $\varGamma_{\lambda}^{\mathcal{X}}$ of $\varGamma$ in a gauge $\mathcal{X}_{\lambda}$ to the representation $\varGamma_{\lambda}^{\mathcal{Y}}$ of $\varGamma$ in a gauge $\mathcal{Y}_{\lambda}$. Now,   following \cite{malik} and using the Baker-Campbell-Haussdorf formula \cite{sopuerta},  one can  generalize eq. (\ref{23}) to write $\Phi_{\lambda}^{*}\varGamma_{\lambda}^{\mathcal{X}}$ in the following way
\begin{equation}
 \Phi_{\lambda}^{*}\varGamma_{\lambda}^{\mathcal{X}}=\exp{\left( \sum_{k=1}^{\infty}\frac{\lambda^{k}}{k!}\mathcal{L}_{\xi_{k}}\right) }\varGamma_{\lambda}^{\mathcal{X}},\label{BCH}
\end{equation}
where $\xi_{k}$ is any vector field on $\mathcal{M}_{\lambda}$.   Substituting eq.(\ref{BCH}) in eq.(\ref{phi}), we have explicitly that
\begin{equation}
 \varGamma_{\lambda}^{\mathcal{Y}}=\varGamma_{\lambda}^{\mathcal{X}}+\lambda\mathcal{L}_{\xi_{1}}\varGamma_{\lambda}^{\mathcal{X}}+\frac{\lambda^{2}}{2}\left(\mathcal{L}_{\xi_{1}}^{2}+\mathcal{L}_{\xi_{2}}\right) \varGamma_{\lambda}^{\mathcal{X}}+\mathcal{O}(\lambda^{3}).\label{expl}
\end{equation}
Replacing  eq.(\ref{a}) and eq.(\ref{b}) into eq.(\ref{expl}), the relations to first and second order perturbations of $\varGamma$ in two different gauge choices are given by
\begin{align}
\delta^{(1)}_{\mathcal{Y}}\varGamma-\delta^{(1)}_{\mathcal{X}}\varGamma & =\mathcal{L}_{\xi_{1}}\varGamma_{0},\label{eq1}\\
\delta^{(2)}_{\mathcal{Y}}\varGamma-\delta^{(2)}_{\mathcal{X}}\varGamma & = 2\mathcal{L}_{\xi_{1}}\delta^{(1)}_{\mathcal{X}}\varGamma_{0}+\left(\mathcal{L}_{\xi_{1}}^{2}+\mathcal{L}_{\xi_{2}}\right)\varGamma_{0},\label{2transform}\end{align}
where the generators of the gauge transformation $\Phi$ are
\begin{equation}
\xi_{1}=Y-X  \quad \text{and} \quad \xi_{2}=\left[ X,Y\right].
\end{equation}
This vector field can be split in their time and space part
\begin{equation}
 \xi^{(r)}_{\mu}\rightarrow\left(\alpha^{(r)},\partial_{i}\beta^{(r)}+d_{i}^{(r)} \right) ,
\end{equation}
here $\alpha^{(r)}$ and $\beta^{(r)}$ are arbitrary scalar functions, and $\partial^{i}d_{i}^{(r)}=0$. The function $\alpha_{(r)}$ determines the choice of constant time hypersurfaces, while $\partial_{i}\beta^{(r)}$ and $d_{i}^{(r)}$  fix the spatial coordinates within these hypersurfaces. The choice of coordinates is arbitrary  and the definitions of perturbations are thus gauge dependent. The gauge transformation given by the eqs. (\ref{eq1}) and (\ref{2transform}) are quite general. To first order  $\varGamma$ is gauge invariant if  $\mathcal{L}_{\xi_{1}}\varGamma_{0}=0$, while to second order one must have other conditions  $\mathcal{L}_{\xi_{1}}\delta^{(1)}_{\mathcal{X}}\varGamma_{0}=\mathcal{L}_{\xi_{1}}^{2}\varGamma_{0}=0$ and $\mathcal{L}_{\xi_{2}}\varGamma_{0}=0$, and so on at high orders.
We will apply the formalism described above to the Robertson-Walker metric, where $k$ does mention  the expansion order. 
\section{FLRW background}\label{background}
At zero order (background), the universe is well des\-cribed by a spatially flat  Friedman-Lema\^{\i}tre-Robertson-Walker metric (FLRW)
\begin{equation}
 ds^{2}=a^{2}(\tau)\left(-d\tau^{2}+ \delta_{ij}dx^{i}dx^{j}\right), 
\end{equation}
with $a(\tau)$ the scale factor with $\tau$ the conformal time. Hereafter the Greek indices run from 0 to 3, and the Latin ones run from 1 to 3 and a prime denotes the derivative with respect to  $\tau$. The Einstein  tensor components in this background are given by
\begin{subequations}
 \begin{eqnarray}
G^{0}_{0}&=&-\frac{3H^{2}}{a^{2}}, \\
 G^{i}_{j}&=&-\frac{1}{a^{2}}\left(2 \frac{a^{\prime\prime}}{a}-H^{2}\right)\delta^{i}_{j},
\end{eqnarray}\end{subequations}
with $H=\frac{a^{\prime}}{a}$ the Hubble parameter. We consider the background  filled with a single barotropic fluid 
where  the  energy momentum tensor is 
\begin{equation}
T_{(fl)\:\nu}^{\mu}=\left(\mu_{(0)}+P_{(0)}\right)u^{\mu}_{(0)}u_{\nu}^{(0)}+P_{(0)}\delta_{\:\nu}^{\mu},\label{tflcero} 
\end{equation}
with $\mu_{(0)}$ the energy density and $P_{(0)}$ the pressure. The comoving observers are defined by the four-velocity $u^{\nu}=(a^{-1},0,0,0)$   with $u^{\nu}u_{\nu}=-1$ and the conservation law for the fluid is 
\begin{equation}
 \mu_{(0)}^{\prime}+3H(\mu_{(0)}+P_{(0)})=0. 
\end{equation}
To deal with the magnetic field, the space-time under study is the fluid permeated by a weak magnetic field,\footnote{With the property  $B_{(0)}^{2}\ll\mu_{(0)}$.} which is a  stochastic field and  can be treated  as a perturbation on the background \cite{2.tsagas,tsagas1}.  
Since the magnetic field has no background contribution, the electromagnetic energy momentum tensor is  automatically gauge invariant at first order (see eq.(\ref{eq1})).
The spatial part of Ohm's law  which is the  proyected current is written by
\begin{equation}
\left(g_{\mu i}+u_{\mu}u_{i}\right)j^{\mu}=\sigma g_{\lambda i}g_{\alpha\mu}F^{\lambda\alpha}u^{\mu},\label{ohm}
\end{equation}
where $j^{\mu}=(\varrho,J^{i})$ is the 4-current and $F^{\lambda\alpha}$ is the electromagnetic tensor given by 
\begin{equation}
F^{\lambda\alpha}=\frac{1}{a^{2}\left(\tau\right)}\left(\begin{array}{cccc}0 & E^{i} & E^{j} & E^{k}\\-E^{i} & 0 & B^{k} & -B^{j}\\-E^{j} & -B^{k} & 0 & B^{i}\\-E^{k} & B^{j} & -B^{i} & 0\end{array}\right).\end{equation}
At zero orden in eq. (\ref{ohm}) the usual Ohms law is found which gives us the relation between the 3-current and the electric field 
\begin{equation}
J_{i}=\sigma E_{i},\end{equation}
where $\sigma$ is the conductivity. Under MHD approximation, large scales the plasma is globally neutral and charge density is neglected ($\varrho=0$) \cite{gio1}.  If the conductivity is infinite ($\sigma\rightarrow\infty$) in the  early universe \cite{ keintaro, *gio2falta}, then eq. (\ref{ohm}) states that the  electric field must vanish ($E_{i}=0$) in order to keep the current density finite \cite{jacson,*landau}. Ho\-wever, the current also should be zero ($J_{i}=0$) because  a nonzero current involves a movement of charge particles that breaks down the isotropy in the background. 
\section{Gauge invariant variables at first order}\label{GI}
We write down the perturbations on a spatially flat Robertson-Walker background. The perturbative expansion at $k-$th order of the matter quantities is given by

\begin{eqnarray}
\mu&=&\mu_{(0)}+\sum_{k=1}^{\infty}\frac{1}{k!}\mu_{(k)},\label{f1dexpansion}\\
B^{2}&=&\sum_{k=1}^{\infty}\frac{1}{k!}B^{2}_{(k)},\label{f2dexpansion}\\
E^{2}&=&\sum_{k=1}^{\infty}\frac{1}{k!}E^{2}_{(k)},\label{f3dexpansion}\\
P&=&P_{(0)}+\sum_{k=1}^{\infty}\frac{1}{k!}P_{(k)},\label{fPexpansion}\\
B^{i}&=&\frac{1}{a^{2}(\tau)}\left(\sum_{k=1}^{\infty}\frac{1}{k!}B_{(k)}^{i}\right),\label{Bexpansion}\\
E^{i}&=&\frac{1}{a^{2}(\tau)}\left(\sum_{k=1}^{\infty}\frac{1}{k!}E_{(k)}^{i}\right),\label{Eexpansion}\\
u^{\mu}&=&\frac{1}{a(\tau)}\left(\delta_{0}^{\mu}+\sum_{k=1}^{\infty}\frac{1}{k!}v_{(k)}^{\mu}\right), \\
j^{\mu}&=&\frac{1}{a(\tau)}\left(\sum_{k=1}^{\infty}\frac{1}{k!}j_{(k)}^{\mu}\right),\end{eqnarray}
where the fields used in above formulas are the average ones (i.e. $B^{2}=\left\langle B^{2}\right\rangle$).\footnote{This happens because the average evolves exactly like $B^{2}$ \cite{buchert}.} We also consider the  perturbations about a FLRW background, so that the metric tensor is given by 
\begin{eqnarray}
g_{00}&=&-a^{2}(\tau)\left(1+2\sum_{k=1}^{\infty}\frac{1}{k!}\psi^{(k)}\right),\\
g_{0i}&=&a^{2}(\tau)\sum_{k=1}^{\infty}\frac{1}{k!}\omega_{i}^{(k)},\\
g_{ij}&=&a^{2}(\tau)\left[\left(1-2\sum_{k=1}^{\infty}\frac{1}{k!}\phi^{(k)}\right)\delta_{ij}\right.\left.+\sum_{k=1}^{\infty}\frac{\chi_{ij}^{(k)}}{k!}\right]. \end{eqnarray}
The perturbations are split into a scalar, transverse vector part, and transverse trace-free tensor
\begin{equation}
\omega_{i}^{(k)}=\partial_{i}\omega^{(k)\Vert}+\omega_{i}^{(k)\bot}, \label{omega1}\end{equation}
with $\partial^{i}\omega_{i}^{(k)\bot}=0$. Similarly we can split $\chi_{ij}^{(k)}$ as
\begin{equation}
\chi_{ij}^{(k)}=D_{ij}\chi^{(k)\Vert}+\partial_{i}\chi_{j}^{(k)\bot}+\partial_{j}\chi_{i}^{(k)\bot}+\chi_{ij}^{(k)\top}, \label{chi1}
\end{equation}
for any tensor quantity.\footnote{With $\partial^{i}\chi_{ij}^{(k)\top}=0$, $\chi_{i}^{(k)i}=0$  and $D_{ij}\equiv\partial_{i}\partial_{j}-\frac{1}{3}\delta_{ij}\partial_{k}\partial^{k}$.} 
Following \cite{1pitrou}, one can find the scalar \textit{gauge invariant variables} at first order given by
\begin{eqnarray}
\Psi^{(1)}&\equiv&\psi^{(1)}+\frac{1}{a}\left(\mathcal{S}_{(1)}^{||}a\right)^{\prime},\\
\Phi^{(1)}&\equiv&\phi^{(1)}+\frac{1}{6}\nabla^{2}\chi^{(1)}-H\mathcal{S}_{(1)}^{||},\\
\Delta^{(1)}&\equiv&\mu_{(1)}+\left(\mu_{(0)}\right)^{\prime}\mathcal{S}_{(1)}^{||},\\
\Delta^{(1)}_{P}&\equiv& P_{(1)}+\left(P_{(0)}\right)^{\prime}\mathcal{S}_{(1)}^{||},
\end{eqnarray}
with $\mathcal{S}_{(1)}^{||}\equiv\left(\omega^{||(1)}-\frac{\left(\chi^{||(1)}\right)^{\prime}}{2}\right)$ the scalar contribution of the shear. The vector modes are
\begin{eqnarray}
{\upsilon}_{(1)}^{i}&\equiv&v_{(1)}^{i}+\left(\chi_{\bot(1)}^{i}\right)^{\prime},\\
\vartheta_{i}^{(1)}&\equiv&\omega_{i}^{(1)}-\left(\chi_{i}^{\bot(1)}\right)^{\prime},\\
\mathcal{V}_{(1)}^{i}&\equiv&\omega_{(1)}^{i}+v_{(1)}^{i}.
\end{eqnarray}
Other gauge invariant variables are  the 3-current,  the charge density and the electric and magnetic fields, because they vanish in the background. The tensor quantities are also gauge invariant because they are null in the background (see eq.(\ref{eq1})).
\subsection{The Ohm  law and the energy momentum tensor}
Using eq. (\ref{ohm})  the Ohm law at first order is
\begin{equation}
J_{i}^{(1)}=\sigma E_{i}^{(1)}. \label{1ohm}\end{equation}
As the conductivity of the medium finite (real MHD), the electric field and the 3-current are nonzero. Now,  the electromagnetic energy momentum tensor is
\begin{eqnarray}
T_{(em)\:0}^{0}&=&-\frac{1}{8\pi} \left( {B}_{(1)}^{2}+{E}_{(1)}^{2}\right),\\
T_{(em)0}^{i}&=&0, \quad T_{(em)i}^{0}=0,\\
T_{(em)\: l}^{i}&=&\frac{1}{4\pi}\left[\frac{1}{6}\left({B}_{(1)}^{2}+{E}_{(1)}^{2}\right) \delta_{\: l}^{i}+\Pi_{\, l(em)}^{i(1)}\right],\end{eqnarray}
where $\Pi_{\, l(em)}^{i(1)}=\frac{1}{3}\left({B}^{2}+{E}^{2}\right)\delta_{\: l}^{i}- B_{\: l}B^{i}-E_{\: l}E^{i}$ is the anisotropic stress tensor  that is gauge invariant by defi\-nition eq.(\ref{eq1}). This term is important to constrain the total magnetic energy because it is source of gravitational waves \cite{Tinaa1}. 
We can see that the electromagnetic energy density appears like a quadratic term in the energy\- momentum tensor, which  means that the  electromagnetic field should be regarded as one half order perturbation.\footnote{Therefore the magnetic field should be split  as $B^{i}=\frac{1}{a(\tau)^{2}}\left(B_{(\frac{1}{2})}^{i}+B_{(1)}^{i}+B_{(\frac{3}{2})}^{i}+....\right)$, see \cite{1/2,*o1/2}.}
Using\- eq. (\ref{tflcero}) and considering the fluctuations of the  matter fields, eqs. (\ref{f1dexpansion}) and (\ref{fPexpansion}), the energy momentum tensor for the fluid is given by
\begin{eqnarray}
T_{(fl)\:0}^{0}&=&-\Delta^{(1)}+\left(\mu_{(0)}\right)^{\prime}\mathcal{S}_{(1)}^{||},\label{tfl1}\\
T_{(fl)\:0}^{i}& =&\left(\mu_{0}+P_{0}\right)\left(\mathcal{V}_{(1)}^{i}-\vartheta_{(1)}^{i}-\left(\chi_{\bot(1)}^{i}\right)^{\prime}\right),\\
T_{(fl)\: i}^{0}&=&-\left(\mu_{0}+P_{0}\right)\mathcal{V}_{i}^{(1)},\\
T_{(fl)\: j}^{i}&=&\left(\Delta_{P}^{(1)}-\left(P_{(0)}\right)^{\prime}\mathcal{S}_{(1)}^{||}\right)\delta_{\: j}^{i}+\Pi_{\: j(fl)}^{i(1)},\label{jtfl1}\end{eqnarray}
where $\Pi_{\: j(f)}^{i(1)}$is the anisotropic stress tensor \cite{durrerlibro}. The above equations are written in terms of gauge invariant variables plus terms as $\mathcal{S}_{(1)}^{||}$ that depend of the gauge choice.
\subsection{The conservation equations}
The total energy momentum conservation equation $\mathcal{T}_{\,\beta;\alpha}^{\alpha}=0$ can be split in  each
component that  is not necessarily conserved independently 
\begin{equation}
\mathcal{T}_{\,\beta;\alpha}^{\alpha}=T_{\,\beta;\alpha}^{\alpha(f)}+T_{\,\beta;\alpha}^{\alpha(E.M)}=0,\label{CEMT}\end{equation}
where
\begin{equation}
T_{\,\beta;\alpha}^{\alpha(E.M)}=F_{\beta\alpha}j^{\alpha}.\end{equation}
Using the eqs. (\ref{tfl1}) and (\ref{jtfl1}), the continuity equation $\mathcal{T}_{\,0;\alpha}^{\alpha}=0$ is given by 
\begin{widetext}
\begin{equation}
\begin{split}
  \left(\Delta^{(1)}\right)^{\prime} & +3H\left(\Delta_{P}^{(1)}+\Delta^{(1)}\right)-3\left(\Phi^{(1)}\right)^{\prime}\left(P_{(0)}+\mu_{(0)}\right)+\left(P_{(0)}+\mu_{(0)}\right)\nabla^{2}\upsilon^{(1)}-3H\left(P_{(0)}+\mu_{(0)}\right)^{\prime}\mathcal{S}_{(1)}^{||}\\
 & -\left(\left(\mu_{(0)}\right)^{\prime}\mathcal{S}_{(1)}^{||}\right)^{\prime}+\left(P_{(0)}+\mu_{(0)}\right)\left(-\frac{1}{2}\nabla^{2}\chi^{(1)}+3H\mathcal{S}_{(1)}^{||}\right)^{\prime} -\left(P_{(0)}+\mu_{(0)}\right)\nabla^{2}\left(\frac{1}{2}\chi^{||(1)}\right)^{\prime}=0. \label{e.c}\end{split}
\end{equation}
The Navier-Stokes equation $\mathcal{T}_{\, i;\alpha}^{\alpha}=0$ is 
\begin{equation}
\left(\mathcal{V}_{i}^{(1)}\right)^{\prime}+\frac{\left(\mu_{(0)}+P_{(0)}\right)^{\prime}}{\left(\mu_{(0)}+P_{(0)}\right)}\mathcal{V}_{i}^{(1)}+4H\mathcal{V}_{i}^{(1)}+\partial_{i}\Psi^{(1)}+\frac{_{\partial_{i}\left(\Delta_{P}^{(1)}-\left(P_{(0)}\right)^{\prime}\mathcal{S}_{(1)}^{||}\right)+\partial_{l}\Pi_{(fl)i}^{(1)l}}}{\left(\mu_{(0)}+P_{(0)}\right)}-\partial_{i}\frac{1}{a}\left(\mathcal{S}_{(1)}^{||}a\right)^{\prime}=0.\label{e.ne}\end{equation}
\end{widetext}
The last equations are written is terms of gauge invariant variables in according to \cite{capri1, *durrerlibro,*nakamuramater}.  It is shown there is not  exist contribution of electromagnetic terms to the conservation equations.  In \cite{20b,*3.}  the energy-momentum tensor of each component is not conserved independently and it is divergence has a source term that takes into account\- the energy and momentum transfer between the components of the photon, electron, proton and the electromagnetic field $T_{\,\beta;\alpha}^{\alpha(f)}=K_{\beta}$.
\section{Maxwell  equations and the cosmological dynamo equation}\label{2sec}
The Maxwell's equations are written as
\begin{equation}
\nabla_{\alpha}F^{\alpha\beta}=j^{\beta}, \quad \nabla_{[\gamma}F_{\alpha\beta]}=0. 
\label{maxwell} \end{equation}
Using eq. (\ref{maxwell}) and the pertubation equations for the metric and electromagnetic fields,  the non-homogeneous Maxwell equations are
\begin{eqnarray}
\partial_{i}E_{(1)}^{i} & = &a\varrho_{(1)},\label{1ME}\\
\epsilon^{ilk}\partial_{l}B^{(1)}_{k} & = &\left(E_{(1)}^{i}\right)^{\prime}
+2HE_{(1)}^{i}+aJ_{(1)}^{i},\label{3ME}\end{eqnarray}
and the homogeneous Maxwell equations
\begin{eqnarray}
B_{k(1)}^{\prime}+2HB_{k}^{(1)}+\epsilon_{\: k}^{ij}\partial_{i}E_{j}^{(1)} & = &0,\label{2ME}\\
\partial^{i}B_{i}^{(1)} & =& 0,\label{4ME}\end{eqnarray}
written also by \cite{brown}. Now using the last equations together  with the ohm's law eq. (\ref{1ohm}), we get an equation which describes  the evolution of magnetic field at first order, this relation is the   
\textit{dynamo equation}:
\begin{widetext}
\begin{equation}
\left(B_{k}^{(1)}\right)^{\prime}+2HB_{k}^{(1)}+\eta\left[\nabla\times\left(\nabla\times B^{(1)}-\left(E^{(1)}\right)^{\prime}-2HE^{(1)}\right)\right]_{k}=0\label{dynamo1},\end{equation}
\end{widetext}
with $\eta=\frac{1}{4\pi\sigma}$ the diffusion coefficient.
The eq. (\ref{dynamo1}) is similar to dynamo equation in MHD but it is in the cosmological context \cite{1,gio1}. This equation has one term  that depends on  $\eta$  which takes into account the dissipation phenomena of the magnetic field (the electric field in this term in general is dropped if we neglect the displacement current). Notice that $\eta$  is a expansion  parameter (due to $\sigma$ is large).  From eq. (\ref{dynamo1}) we see  that for  finite $\eta$,  the diffusion term should not be neglected. Care should be taken the assumption $\eta=0$, because it could break at small scales \cite{gio2falta}. In the frozen in condition of magnetic field lines,  where amplification of the field is not taking account, the last equation has the solution $\textbf{B}=\frac{\textbf{B}_{0}}{a^{2}(\tau)}$ where $\textbf{B}_{0}$ is the actual magnetic field, the actual value of the scale factor $a_{0}(\tau)=1$ and $\textbf{B}$ is the magnetic field when the scale factor was $a(\tau)$. 
\section{Generalization at second order} \label{secapedb}
Following  \cite{nakamuranuevo}  the variable $\delta^{(2)}\boldsymbol{T}$  defined by
\begin{equation}
\delta^{(2)}_{\mathcal{X}}\boldsymbol{T}\equiv\delta^{(2)}_{\mathcal{X}}\varGamma-2L_{X}\left(\delta^{(1)}_{\mathcal{X}}\varGamma\right)+L_{X}^{2}\varGamma_{0};
\end{equation}
is introduced. Inspecting the gauge transformation eq. (\ref{2transform}) one can see that $\delta^{(2)}\boldsymbol{T}$ is transformed as 
\begin{equation}
\delta^{(2)}_{\mathcal{Y}}\boldsymbol{T}-\delta^{(2)}_{\mathcal{X}}\boldsymbol{T}=L_{\sigma}\varGamma_{0},\label{eq:sigma}\end{equation}
with $\sigma=\xi_{2}+\left[\xi_{1},X\right]$ and  $X$ is the gauge dependence part in linear order perturbation.
The gauge transformation rule eq. (\ref{eq:sigma}) is identical to the  gauge transformation at linear order eq. (\ref{eq1}). 
This property is  general and is the key to extend this theory to second order
\begin{equation}
L\left[\delta^{2}\boldsymbol{T}\right]=S\left[\delta\boldsymbol{T},\delta\boldsymbol{T}\right].
 \end{equation}
Notice that first and second order equations are similar, however the last  have as sources the coupling between li\-near perturbations variables. Using eqs. (\ref{2transform}) and (\ref{eq:sigma}) we  arrive to the gauge invariant quantities at second order. 
This coupling appearing  as the quadratic terms of the linear perturbation is due to the nonlinear effects of the Einstein field equations, besides one can classify them again in scalar, vector and tensor modes where this modes couple with each other. Now, to clarify the physical behaviors of perturbations at this order we should obtain the gauge invariant quantities and express these equations of movements in terms of these quantities.    

The scalar gauge invariants are given by
\begin{eqnarray}
\Psi^{(2)}&\equiv&\psi^{(2)}+\frac{1}{a}\left(\mathcal{S}_{(2)}^{||}a\right)^{\prime}+\mathcal{T}^{1}(\mathcal{O}^{(2)})\,, \label{psi2ordergi} 
\\
\Phi^{(2)}&\equiv&\phi^{(2)}+\frac{1}{6}\nabla^{2}\chi^{(2)}-H\mathcal{S}_{(2)}^{||}+\mathcal{T}^{2}(\mathcal{O}^{(2)}) \,,
\\
\Delta_{\mu}^{(2)}&\equiv&\mu_{(2)}+\left(\mu_{(0)}\right)^{\prime}\mathcal{S}_{(2)}^{||}+\mathcal{T}^{3}(\mathcal{O}^{(2)})\,, \\
\Delta_{\varrho}^{(2)}&\equiv&\varrho^{(2)}+\mathcal{T}^{4}(\mathcal{O}^{(2)})\,, \\
\Delta_{B}^{(2)}&\equiv&B^{2}_{(2)}+\mathcal{T}^{4}(\mathcal{O}^{(2)})\,, \\
\Delta_{E}^{(2)}&\equiv&E^{2}_{(2)}+\mathcal{T}^{6}(\mathcal{O}^{(2)})\,, \\
\upsilon^{(2)}&\equiv&v^{(2)}+\left(\frac{1}{2}\chi^{||(2)}\right)^{\prime}+\mathcal{T}^{7}(\mathcal{O}^{(2)})\,, \end{eqnarray}
with $\mathcal{S}_{(2)}^{||}\equiv\left(\omega^{||(2)}-\frac{\left(\chi^{||(2)}\right)^{\prime}}{2}\right)+\mathcal{T}^{8}(\mathcal{O}^{(2)})$. The expre\-ssion for $\mathcal{T}^{8}(\mathcal{O}^{(2)})$ is given in  appendix \ref{app_evol}. In this case $\mathcal{S}_{(2)}^{||}$ can be interpreted like shear at second order. Again it is showed that it is similar to found at first order but it has a source term which is quadratic in the first order fuctions of the transformations.
The vector modes found are as follows
\begin{eqnarray}
\upsilon_{(2)}^{i}&\equiv&v_{(2)}^{i}+\left(\chi_{\bot(2)}^{i}\right)^{\prime}+\mathcal{T}^{9}(\mathcal{O}^{(2)}),\\
\vartheta_{i}^{(2)}&\equiv&\omega_{i}^{(2)}-\left(\chi_{i}^{\bot(2)}\right)^{\prime}+\mathcal{T}^{10}(\mathcal{O}^{(2)}),\\
\mathcal{V}_{(2)}^{i}&\equiv&\omega_{(2)}^{i}+v_{(2)}^{i}+\mathcal{T}^{11}(\mathcal{O}^{(2)}),\\
\Pi_{ij}^{(2)T}&\equiv&\Pi_{ij}^{(2)fl}+\Pi_{ij}^{(2)em}+\mathcal{T}^{13}(\mathcal{O}^{(2)}),\label{pi2ordergi}
\end{eqnarray}
The electromagnetic fields modes (from $F^{\lambda\alpha}$) are then given by
\begin{eqnarray}
\mathcal{E}^{(2)}_{i} &=& E^{(2)}_{i}+2\left[\frac{1}{a^{2}}\left(a^{2}E^{(1)}_{i} \alpha^{(1)}\right)^{\prime}+\left(\xi^{\prime}_{(1)}\times B^{(1)}\right)_{i} \right. \nonumber \\
&+& \left. \xi^{l}_{(1)}\partial_{l}E_{i}^{(1)}+E_{l}^{(1)}\partial_{i}\xi^{l}_{(1)}\right]   \,,
\\
\mathcal{B}^{(2)}_{i} &=& B^{(2)}_{i}+2\left[\frac{\alpha^{(1)}}{a^{2}}\left(a^{2}B^{(1)}_{i} \right)^{\prime}+\xi^{l}_{(1)}\partial_{l}B^{(1)}_{i} \right. \nonumber 
\\
&+& \left. B_{i}^{(1)}\partial_{l}\xi^{l}_{(1)}+\left(E^{(1)}\times \nabla\alpha^{(1)}\right)_{i}-B_{l}^{(1)}\partial^{l}\xi^{(1)}_{i}\right]  \,,
\\
\varrho^{(2)}_{(Inv.)} &=& \varrho^{(2)}+2\left[\left(\varrho_{(1)}^{\prime}-H\varrho^{(1)}\right)\alpha^{(1)}+\xi^{i}_{(1)}\partial_{i}\varrho^{(1)} \right. \nonumber
\\
&-& \left. \alpha^{\prime}_{(1)}\varrho^{(1)}-J^{i}_{(1)}\partial_{i}\alpha^{(1)}\right] \,,
\\
\mathcal{J}_{(2)}^{i} &=& J_{(2)}^{i}+2\left[\left( \mathcal({J}_{(1)}^{i})^{\prime}-H\mathcal{J}_{(1)}^{i}\right)\alpha^{(1)}+\xi^{l}_{(1)}\partial_{l}J^{i}_{(1)} \right. \nonumber \\
&-& \left. \varrho^{(1)}(\xi^{i}_{(1)})^{\prime}-J^{l}_{(1)}\partial_{l}\xi^{i}\right] \,,\end{eqnarray}
which are gauge invariant quantities  for electromagnetic fields. All these variables are similar to the quantities obtained  at first orden, but in second order case appear  as sources as ${T}^{k}(\mathcal{O}^{(2)})$ that depend of the gauge choice and the coupling with terms of first order. The explicit calculation of 
$\mathcal{T}^{k}(\mathcal{O}^{(2)})$ is shown in \cite{mollerach,1pitrou}. 
\subsection{The Ohm law and the  energy momentum tensor}
Using eqs. (\ref{ohm}), (\ref{Bexpansion}) and (\ref{Eexpansion}), we get the Ohm law at second order 
\begin{eqnarray}
\mathcal{J}_{i}^{(2)} &&= 4J_{i}^{(1)}\Phi^{(1)}+\boldsymbol{S}_{i}^{1}(\mathcal{O}^{(2)}) \nonumber \\
&& +\varrho^{(1)}\upsilon_{i}^{(1)}+2 \sigma
\left(\left(\mathcal{V}_{(1)}\times B^{(1)}\right)_{i}+\frac{1}{2}\mathcal{E}^{(2)}_{i}\right. \nonumber \\
&& \left. -2E_{i}^{(1)}\left(\Phi^{(1)}-\frac{1}{2}\Psi^{(1)}\right)+\boldsymbol{S}_{i}^{2}(\mathcal{O}^{(2)})\right)\,. \label{ohm2}\end{eqnarray}

In this case we see that 3-current has a  type of Lorentz term and shows coupling between first order terms that  affect the evolution of the current. Hereafter  the functions $\boldsymbol{S}_{i}^{n}(\mathcal{O}^{(2)})$ with $n\in \mathbb{Z}$ and $i$ being the component, gives us the gauge dependence.
The last equation shows also a coupling between the electric field and terms like ($\Phi^{(1)}-\frac{1}{2}\Psi^{(1)}$) that  is associated to tidal forces (this quantity is similar to scalar part of the electric part of Weyl tensor) and the first right hand term between the cu\-rrent and perturbation in the curvature. There exist mo\-dels where the coupling of the charge particles and the field is important for explaining some phenomena like collapse or generation of magnetic field during recombination period. In this case the Ohm law shown in eq.(\ref{ohm2}) should be generalized and  terms like Biermann battery and Hall effect should appear.      
 Doing the expansion at second order in the fluid energy momentum tensor, one finds the following expressions
\begin{eqnarray}
T_{(2)\:0}^{0}&=&-\frac{\Delta_{\mu}^{(2)}}{2}-\left(\mu_{(0)}+P_{(0)}\right)\left(\upsilon_{l}^{(1)}\upsilon_{(1)}^{l}+\vartheta_{l}^{(1)}\upsilon_{(1)}^{l}\right)\nonumber\\&+&\boldsymbol{S}^{3}(\mathcal{O}^{(2)}),\\
T_{(2)\:0}^{i}& =&-\left(\mu_{(0)}+P_{(0)}\right)\left(\frac{\mathcal{V}_{(2)}^{i}-\vartheta_{(2)}^{i}}{2}+\Psi^{(1)}\upsilon_{(1)}^{i}\right)\nonumber\\
&-&\left(\Delta_{\mu}^{(1)}+\Delta_{P}^{(1)}\right)\upsilon_{(1)}^{i}+\boldsymbol{S}_{4}^{i}\left(\mathcal{O}^{(2)}\right),\\
T_{(2)\:i}^{0}&=&-\left(\mu_{(0)}+P_{(0)}\right)\left(\frac{\mathcal{V}_{i}^{(2)}}{2}-2\vartheta_{i}^{(2)}\Psi^{(1)}-2\upsilon_{i}^{(1)}\Phi^{(1)}\right.\nonumber\\
&+&\left.\upsilon_{(1)}^{j}\chi_{ij}^{(1)}-\upsilon_{i}^{(1)}\Psi^{(1)}\right)-\left(\Delta_{\mu}^{(1)}+\Delta_{P}^{(1)}\right)\mathcal{V}_{i}^{(1)}\nonumber\\&+&\boldsymbol{S}_{i}^{5}\left(\mathcal{O}^{(2)}\right),\\
T_{(2)\: j}^{i}&=&\frac{1}{2}\Delta_{P}^{(2)}\delta_{\: j}^{i}+\frac{1}{2}\Pi_{\:j}^{i(2)}+\boldsymbol{S}^{6i}_{j}\left(\mathcal{O}^{(2)}\right)\nonumber\\
&+&\left(\mu_{(0)}+P_{(0)}\right)\left(\upsilon_{j}^{(1)}\upsilon_{(1)}^{i}+\vartheta_{j}^{(1)}\upsilon_{(1)}^{i}\right),
\end{eqnarray}
similar to \cite{nakamuramater}.  Now  consider eq. (\ref{2ME}) the electromagnetic momentum tensor at second order is
\begin{eqnarray}
T_{(em)\:0}^{0}&=&-\frac{1}{8\pi}\left(\Delta_{E}^{(2)}+\Delta_{B}^{(2)}+\boldsymbol{S}^{8}(\mathcal{O}^{(2)})\right),\\
T_{(em)0}^{i}&=&\frac{1}{4\pi}\left[\left.-\epsilon^{ikm}E_{k}^{(1)}B_{(1)}^{m}+\boldsymbol{S_{9}^{i}}(\mathcal{O}^{(2)})\right]\right.,\\
T_{(em)i}^{0}&=&\frac{1}{4\pi}\left[\epsilon_{i}^{\, km}E_{k}^{(1)}B_{(1)}^{m}+\boldsymbol{S}_{10i}(\mathcal{O}^{(2)})\right],\\
T_{(em)\:l}^{i}&=&\frac{1}{4\pi}\left[\frac{1}{6}\left(\Delta_{E}^{(2)}+\Delta_{B}^{(2)}+\boldsymbol{S}_{4l}^{i}(\mathcal{O}^{(1)})\right)\delta_{\: l}^{i}\right. \nonumber\\
&+&\left.\Pi_{\, l(em)}^{i(2)}+\boldsymbol{S}_{11l}^{i}\left(\mathcal{O}^{(2)}\right)\right]. 
\end{eqnarray}
Using eq. (\ref{CEMT}) the continuity equation is given by
\begin{eqnarray}
\left(\Delta_{\mu}^{(2)}\right)^{\prime} &+&3H\left(\Delta_{P}^{(2)}+\Delta_{\mu}^{(2)}\right)-3\left(\Phi^{(2)}\right)^{\prime}\left(P_{(0)}+\mu_{(0)}\right)\nonumber\\
&+&\left(P_{(0)}+\mu_{(0)}\right)\nabla^{2}\upsilon^{(2)}=-a^{4}\left(2E_{i}^{(1)}J_{(1)}^{i}\right)\nonumber\\
&-&\boldsymbol{S}_{12}\left(\mathcal{O}^{(2)}\right),\label{energy2}\end{eqnarray}
and  the Navier-stokes equation
\begin{eqnarray}
&&\frac{1}{2}\frac{\left[\mu_{(0)}\left( 1+w \right)\mathcal{V}_{i}^{(2)}\right]^{\prime}}{\mu_{(0)}\left(1+w \right)}+2H\mathcal{V}_{i}^{(2)}+\frac{1}{2}\frac{\partial_{i}P^{(2)}+2\partial_{j}\Pi_{i}^{j(2)}}{\mu_{(0)}\left(1+w \right)} \nonumber \\
&+&\frac{1}{2}\partial_{i}\Psi^{(2)}+\boldsymbol{S}^{13}_{i}\left(\mathcal{O}^{(2)}\right)= \frac{a^{4}\left(E_{i}^{(1)}\varrho_{(1)}+\epsilon_{ijk}J_{(1)}^{j}B_{k}^{(1)}\right)}{\mu_{(0)}\left(1+w\right)},\label{navierstokes2} \nonumber \\ \end{eqnarray}
  where $w=\frac{P_{(0)}}{\mu_{(0)}}$ and  $\boldsymbol{S}^{13}_{i}$ is shown  in  appendix \ref{app_evol2}.  Therefore, electromagnetic fields affect the evolution of  matter energy density $\Delta_{\mu}^{(2)}$ and the peculiar velocity $\mathcal{V}_{i}^{(2)}$ also, these fields  influence the large structure formation and can leave imprints on the temperature anisotropy pattern of the CMB \cite{capri1,*Tinaf,*gio3,*gio4}.
\section{The Maxwell equations and the cosmological dynamo at second order}
\label{sec:3}
Using the eq. (\ref{1ohm}), the non homogeneous  Maxwell's equations are
\begin{eqnarray}
\partial_{i}\mathcal{E}_{(2)}^{i}&=&-4E_{(1)}^{i}\partial_{i}\left(\Psi^{(1)}-3\Phi^{(1)}\right)+a\Delta_{\varrho}^{(2)}\nonumber\\
&-&\boldsymbol{S}_{14}\left(\mathcal{O}^{(2)}\right),\\
\left(\nabla\times\mathcal{B}^{(2)}\right)^{i}&=&2E_{(1)}^{i}\left(2\left(\Psi^{(1)}\right)^{\prime}-6\left(\Phi^{(1)}\right)^{\prime}\right)+\left(\mathcal{E}_{(2)}^{i}\right)^{\prime}\nonumber\\
&+&2H\mathcal{E}_{(2)}^{i}+2\left(2\Psi^{(1)}-6\Phi^{(1)}\right)\left(\nabla\times B_{(1)}\right)^{i}\nonumber\\
&+&a\mathcal{J}_{(2)}^{i}+\boldsymbol{S}^{i}_{15}\left(\mathcal{O}^{(2)}\right).\end{eqnarray}
While the homogeneous Maxwell's equations are
\begin{eqnarray}
\frac{1}{a^{2}}\left(a^{2}\mathcal{B}_{k}^{(2)}\right)^{\prime}& + & \left(\nabla\times\mathcal{E}_{j(2)}\right)_{k}=-\boldsymbol{S}_{k}^{17}\left(\mathcal{O}^{(2)}\right)\,, \\
\partial_{i}\mathcal{B}^{i(2)}&=& 0\,.
\end{eqnarray}
Again the $\boldsymbol{S}_{k}^{n}$ terms carry out the gauge dependence. Using\-  the Maxwell equations together with the Ohm law at second order and following the same methodology for the first order case, we get the cosmological dynamo equation that  describes the evolution of the magnetic field at second order
\begin{widetext}\begin{align}
\left(\mathcal{B}_{k}^{(2)}\right)^{\prime} & +2H\left(\mathcal{B}_{k}^{(2)}\right)+\eta\Biggl[\nabla\times\Bigl(\frac{1}{a}\left(\left(\nabla\times\mathcal{B}^{(2)}\right)-2E_{(1)}\left(2\left(\Psi^{(1)}\right)^{\prime}-6\left(\Phi^{(1)}\right)^{\prime}\right)\right.\nonumber\\
&\left.-\left(\mathcal{E}_{(2)}\right)^{\prime}-2H\mathcal{E}_{(2)}-2\left(\nabla\times B_{(1)}\left(2\Psi^{(1)}-6\Phi^{(1)}\right)\right)-\boldsymbol{S}_{15}\left(\mathcal{O}^{(1)}\right)\right) -\varrho^{(1)}\upsilon^{(1)}+\boldsymbol{S}^{1}\left(\mathcal{O}^{(2)}\right)\biggr)\Biggr]_{k}\nonumber\\
& +\Biggl(\nabla\times\left[-2\left(\mathcal{V}_{(1)}\times B^{(1)}\right)-2E^{(1)}\Psi^{(1)}-2\boldsymbol{S}^{2}\left(\mathcal{O}^{(1)}\right) \right]\Biggr)_{k} =-\boldsymbol{S}_{k}^{17}\left(\mathcal{O}^{(2)}\right),\label{2dynamoequation}
\end{align}\end{widetext}
where the value of $\varrho^{(1)}$ can be found to resolve the di\-fferential equation given in appendix \ref{app_evol2}. Thus the pertur\-bations in the space-time play an important role in the evolution of primordial magnetic fields. The eqs. (\ref{dynamo1}) and (\ref{2dynamoequation})  are  dependent on geometrical quantities (perturbation in the gravitational potential, curvature, velocity ...).  These quantities evolve  according to the Einstein field equations (the Einstein field equation to second order are given in \cite{nak}). In this way, the equation Eq. (\ref{2dynamoequation}) tells us how the magnetic field evolves according to the scale of the perturbation. In  sub-horizon scale, the contrast density and the geometrical quantities grow. Hence, the dynamo term should amplify the magnetic field.
As  a final comment  we point out that in order to solve the dynamo like-equation for the magnetic field is necessary  to solve the Einstein field equations to the second order together with the conservation equations.
\section{Specifying to Poisson gauge}
It is possible to fix the four degrees of freedom by imposing gauge conditions.
 If we impose  the gauge restrictions
 \begin{equation}
 \partial^{i}\omega_{i}^{(r)}=\partial^{i}\chi_{ij}^{(r)}=0,
\end{equation}
all equation can be written in terms in quantities independent of the coordinates\cite{Bert}. This gauge is called \textit{Poisson gauge} and it is the gravitational analogue of Coulomb gauge in electromagnetism (see appendix \ref{app_evol}). The perturbed metric in the Poisson gauge reads
\begin{eqnarray}
g_{00}&=&-a^{2}(\tau)\left(1+2\psi^{(1)}+\psi^{(2)}\right)\,, \label{metricpoison}\nonumber\\
g_{ij}&=&a^{2}(\tau)\left[\left(1-2\phi^{(1)}-\phi^{(2)}\right)\delta_{ij}+\frac{2\chi_{ij}^{(1)\top}+\chi_{ij}^{(2)\top}}{2}\right]\,, \nonumber\\ 
g_{0i}&=&a^{2}(\tau)\left(\omega_{i}^{(1)\bot}+\frac{\omega_{i}^{(2)\bot}}{2}\right)\,, 
\end{eqnarray}
where $\omega^{\parallel}, \chi^{\parallel}, \chi_{i}^{\bot}$ are null. In this case the dynamo equation in the Poisson gauge is given by
\begin{widetext}\begin{eqnarray}
B_{k(2)}^{\prime}&+&2HB_{k}^{(2)}+\eta \left[  \nabla \times \left(\nabla \times B_{(2)}\right)-\left(\nabla\times E^{\prime}_{(2)}  \right)-2H\left(\nabla \times E^{(2)} \right)-4\left(\Psi^{\prime}_{(1)}-3\Phi^{\prime}_{(1)} \right)\left( \nabla \times E_{(1)}\right) \right. \nonumber\\  
&-&4\nabla\left(\Psi^{\prime}_{(1)}-3\Phi^{\prime}_{(1)}\right)\times E^{(1)}+4\left(\nabla\times\left( \nabla\left(\Psi^{(1)}-3\Phi^{(1)}\right)\times B^{(1)}\right)  \right)-4\left(\left(  \nabla\times\left(\nabla\times B^{(1)}\right)  \right. \right.         \nonumber \\
&-& \left. \left. \nabla\times E^{\prime}_{(1)}-2H\left( \nabla \times E^{(1)}\right)\right) \Phi^{(1)} + \nabla\Phi^{(1)}\times \left(\nabla \times B^{(1)}-E^{\prime}_{(1)}-2HE_{(1)} \right)\right) -\left(\nabla \varrho^{(1)} \right)\times v^{(1)}\nonumber \\    
&+& \left. 2 \nabla \times \left( \left( \nabla \times B^{(1)}-\frac{1}{a^{2}}\left(a^{2}E_{(1)}\right)^{\prime}\right)\cdot\chi^{\top}_{(1)}\right)-\varrho^{(1)}\left(\nabla \times v^{(1)}\right)\right]_{k}-2\left( \nabla \times \left( \left(v^{(1)}+\omega^{\bot}_{(1)}\right)\times B^{(1)}\right)\right)_{k} \nonumber \\
&+&4\left( \left( \nabla \left(\Phi^{(1)}-\frac{\Psi^{(1)}}{2}\right) \times E^{(1)}\right)+\left(\Phi^{(1)}-\frac{\Psi^{(1)}}{2} \right)\left(\nabla \times E^{(1)}\right)\right)_{k}-2\nabla\times \left(E^{(1)}\cdot \chi^{\top}_{(1)}\right)_{k}=0, 
\end{eqnarray}\end{widetext}
where $E^{(1)}\cdot \chi^{\top}_{(1)}=E^{(1)}_{i}\chi^{ij}_{\top(1)}$.  The last equation is a specific case of the  equation Eq. (\ref{2dynamoequation}) where we fix the gauge (coordinate fixing). It is important to notice the re\-levance of the geometrical perturbation quantities in the evolution of the magnetic fields, again we see the influence of the tidal and Lorentz forces in the amplification of the fields. 
In some sense, the above equation differs  from equation Eq. (\ref{2dynamoequation}) due to the fact we fix the adequately choice of the perturbation functions (we choose a gauge for writing the equation of motion without the presence of unphysical modes) while before we just wrote the equations in terms of gauge invariant quatities which were built up  with the formalim explained in the fist sections, plus terms which have in taken into account the dependence of the gauge and where we need  to fix them. 
\section{Weakly magnetized FLRW-background}\label{2part}
In this section we work a magnetized FLRW, i.e   we  
allow the presence of a weak magnetic field into our FLRW background  with the  property $B_{(0)}^{2}\ll\mu_{(0)}$ which  must to be sufficiently random to satisfy $\left\langle B_{i} \right\rangle=0$ and $ \left\langle B_{(0)}^{2}\right\rangle =\left\langle B^{(0)}_{i}B_{(0)}^{i}\right\rangle\neq0$ to ensure that  symmetries and the evolution of the background  remain unaffected.
Again we work under MHD approximation, and thus in large scales the plasma is globally neutral, charge density is neglected and the electric field with the current  should be zero, thus the only zero order magnetic variable is $B_{(0)}^{2}$ \cite{tsagas1}. The evolution of  density magnetic field can be found contracting the induction equation with  $B_{i}$ arriving at
\begin{equation}
 \left( B_{(0)}^{2}\right)^{\prime }=-4HB_{(0)}^{2},
\end{equation}
showing $B^{2}\sim a^{-4}$ in the background.  Bianchi models are often used to describe the presence of a magnetic field in the universe due to anisotropic properties of this metric.  However, as we are dealing with  weak magnetic fields, it is worth to assuming the presence of a magnetic field in a FLRW metric as background.
Indeed, the authors in \cite{tsagasroy} found that, although there is a profound distinction between  the Bianchi I equations and the FLRW approxi\-mation, at the weak field limit, these differences are reduced dramatically, and therefore the linearised Bianchi equations are the same as with the FLRW ones.  Under these conditions, we find that to zero order the electromagnetic energy momentum tensor in the background is given by:
 \begin{eqnarray}
T_{(em)\:0}^{0} & =&-\frac{1}{8\pi}B_{(0)}^{2},\label{temcero1}\\
T_{(em)\: i}^{0} & =&T_{(em)\:0}^{i}=0,\\
T_{(em)\: l}^{i} & =&\frac{1}{24\pi}B_{(0)}^{2}\delta_{\: l}^{i}.\label{temcero}\end{eqnarray}
The magnetic anisotropic stress is treated as a first-order perturbation due to stochastic properties of the field, therefore it does not contribute to the above equations.
 We can see in eqs. (\ref{tflcero}) and (\ref{temcero1})-(\ref{temcero}), that  fluid  and  electromagnetic  energy-momentum tensor are diagonal tensors, that is, are consistent with the condition of an isotropic and homogeneous background \cite{tsagas1}.
If we consider the average magnetic density of the background di\-fferent to zero, the perturbative expansion at $k-$th order of the magnetic density is given by
 \begin{equation}
B^{2}=B_{(0)}^{2}+\sum_{k=1}^{\infty}\frac{1}{k!}B^{2}_{(k)},\label{1fdexpansion}
\end{equation} 
where at first order we get a gauge invariant term which describes the magnetic energy density
\begin{equation}
\Delta^{(1)}_{mag}\equiv\ B_{(1)}^{2}+\left(B_{(0)}^{2}\right)^{\prime}\mathcal{S}_{(1)}^{||};\end{equation}
one can find that average density of the background field decays as $B_{(0)}^{2}\sim\frac{1}{a^{4}(\tau)}$ \cite{Subramanian}. At first orden we work with finite conductivity (real MHD), in this case the electric field and the current becomes nonzero, therefore using the eq. (\ref{ohm}) and assuming the ohmic current is not neglected, we find the Ohm's law  
\begin{equation}
J_{i}^{(1)}=\sigma\left[  E_{i}^{(1)}+\left(\mathcal{V}^{(1)}\times B^{(0)}\right)_{i}\right]. \label{11ohm}\end{equation}
In the last  equation the  Lorentz force appears at first order when a magnetic field is consider as a part of  the background. 
Again doing the  same procedure  described before, but taking a weak magnetic field as a contribution from the background we shall show the implication of this supposition afterword. 
The electromagnetic energy momentum tensor at first order is given by 
\begin{eqnarray}
T_{(em)\:0}^{0}&=&-\frac{1}{16\pi} {F}_{(1)}^{2},\\
T_{(em)0}^{i}&=&\frac{1}{4\pi}\left[B_{(0)}^{2}\vartheta_{(1)}^{i}\right.\nonumber\\
&-&\left.\epsilon^{ikm}E_{k}^{(1)}B_{(0)}^{m}+B_{(0)}^{2}\left(\chi_{\bot(1)}^{i}\right)^{\prime}\right]\\
T_{(em)i}^{0}&=&\frac{1}{4\pi}\left[\epsilon_{i}^{\, km}E_{k}^{(1)}B_{(0)}^{m}\right],\\
T_{(em)\: l}^{i}&=&\frac{1}{4\pi}\left[\frac{1}{12}{F}_{(1)}^{2}\delta_{\: l}^{i}+\Pi_{\, l(em)}^{i(1)}\right],\end{eqnarray}
where
\begin{eqnarray}
{F}_{(1)}^{2}&=&2\Delta_{(mag)}^{(1)}-8\Phi^{(1)}B_{(0)}^{2}-2\left(B_{(0)}^{2}\right)^{\prime}\mathcal{S}_{(1)}^{||}\nonumber\\
&+&\frac{4}{3}\nabla^{2}\chi^{(1)}B_{(0)}^{2}-8H\mathcal{S}_{(1)}^{||}B_{(0)}^{2},\end{eqnarray}
and $\Pi_{\, l(em)}^{i(1)}=\frac{1}{3}\left(\Delta^{(1)}_{mag}+{E}^{2}\right)\delta_{\: l}^{i}- B_{\: l}B^{i}-E_{\: l}E^{i}$ is the anisotropic stress that  appears as a perturbation of the background, this term is important to constraining the total magnetic energy because it is a source of gravitational waves \cite{Tinaa1}. 
The above equations are written in terms of gauge invariant variables plus terms as $\mathcal{S}_{(1)}$ which are gauge dependent.
Now, using the above equations eqs. (\ref{1ME}),(\ref{2ME}),(\ref{3ME}) and (\ref{4ME})  with the ohm's law eq. (\ref{11ohm}), we arrive to the dynamo equation that gives us the evolution of magnetic field to first order
\begin{widetext}
\begin{equation}
\left(B_{k}^{(1)}\right)^{\prime}+2HB_{k}^{(1)}+\eta\left[\nabla\times\left(\nabla\times B^{(1)}-\left(E^{(1)}\right)^{\prime}-2HE^{(1)}\right)\right]_{k}+\left(\nabla\times \left( B_{(0)}\times\mathcal{V}_{(1)}\right) \right)_{k}=0\label{dynamo11}.\end{equation}\end{widetext}
When we suppose a weak magnetic field on the background, in the  dynamo equation  a new term  called \textit{dynamo term} appears which could  amplify the magnetic field. This term depends of the evolution in $\mathcal{V}_{(1)}$, see eq. (\ref{e.ne}), and also from eq. (\ref{e.ne}), it seems likely when matter and velocity perturbation grow the  dynamo term amplifies the magnetic field, this is a difference with the first approach where the dynamo term just appears  at second order. 
For convenience it is better use the Lagrangian coordinates which are comoving with the local Hubble  flow. So we use the  convective derivative which is evaluated according to the operator formula  (i.e $ \dfrac{d}{dt}=\dfrac{\partial}{\partial t}+\mathcal{V}_{(1)}^{i}\partial_{i} $). In this picture the magnetic field lines are frozen into the fluid. Using the well known identity formula 
\begin{equation}
 \nabla\times\left(a\times b\right)=a\left(\nabla\cdot b \right)-b\left(\nabla\cdot a\right)+\left(b\cdot\nabla\right)a-\left(a\cdot\nabla\right)b,  
\end{equation}
we obtain the following result 
\begin{equation}
\dfrac{dB_{i}}{dt}+2HB_{i}=B_{j}\left(\dfrac{\partial \mathcal{V}^{(1)}_{i}}{\partial x_{j}}-\frac{1}{3}\delta_{ij}\dfrac{\partial 
\mathcal{V}^{(1)}_{k}}{\partial x_{k}}\right)+\frac{2}{3}B_{i} \dfrac{\partial\mathcal{V}^{(1)}_{j}}{\partial x_{j}}, 
\label{formulasinshear}\end{equation}
where diffusion term will not be considered for the moment. The first term in the right hand is associated with the shear and the last term describes the expansion of the region where $\mathcal{V}^{(1)}$ is not zero.  In the case of a homogeneous collapse, $B \sim \mathcal{V}^{-\frac{2}{3}}$ gives rise to amplification of  the magnetic field in places where gravitational collapse takes place.  Now we  write eq.(\ref{dynamo11}) in the Poisson gauge  getting the following 
\begin{widetext}
\begin{eqnarray}
\frac{dB^{k}_{(1)}}{dt} &+&  2HB_{(1)}^{k}+\eta \left[-\nabla^{2} B^{k}_{(1)}-\left( \nabla \times \left(\frac{1}{a^{2}} \frac{d(a^{2}E_{(1)})}{dt}-\mathcal{V}_{(1)}^{i}\partial_{i}E_{(1)}\right)\right)^{k}  -B_{(0)}^{k}\nabla^{2}\left(\Psi^{(1)}-3\Phi^{(1)}\right)\right. \nonumber \\
&+& \left. \left( B^{(0)}\cdot \nabla\right) \partial^{k} \left( \Psi^{(1)}-3 \Phi^{(1)} \right)-\left( \nabla\left(\Psi^{(1)}-3 \Phi^{(1)} \right) \cdot \nabla \right)B^{k}_{(0)} \right] =B_{l}^{(0)}\sigma_{(1)}^{lk}-\frac{2}{3}B^{k}_{(0)}\partial_{l} \mathcal{V}_{(1)}^{l} \label{2ordendinapoison}\,,
\end{eqnarray}\end{widetext}
where $\sigma^{lk}_{(1)}$ is the shear found in eq.(\ref{formulasinshear}). The last term on the left-hand side in eq. (\ref{2ordendinapoison}) should vanish due to the  background isotropy.  The evolution of magnetic field following the last equation is highly dependent of term $\Psi^{(1)}-3\Phi^{(1)}$. If the perturbations are turned off, one can check that last equation recovers to the dynamo equation found in the literature.  It should be noted terms as $\left\langle B^{k}_{(0)}\right\rangle$ are zero due to statistical  field properties,  therefore  contracting  eq. \ref{2ordendinapoison}  with magnetic field $B_{k}^{(1)}$, we arrive at an  equation at second order which we can  physi\-cally study the evolution of the density magnetic field
\begin{widetext}
\begin{eqnarray}
&&\frac{d\Delta_{(mag)}^{(2)}}{dt} + 4H\Delta_{(mag)}^{(2)}+2\eta \left[-B^{(1)}\cdot \nabla^{2}B^{(1)}-B^{(1)}\cdot\left(\nabla \times \left(\frac{1}{a^{2}} \frac{d(a^{2}E_{(1)})}{dt}-\mathcal{V}_{(1)}^{i}\partial_{i}E_{(1)}\right)\right)  \right. \label{21dynamoequationa} \\
&-& \left.\frac{1}{2}\Delta_{(mag)}^{(1)}\nabla^{2}\left( \Psi^{(1)}-3 \Phi^{(1)} \right) +B^{k}_{(1)}\left(B^{(0)}\cdot \nabla \right)\partial_{k}\left( \Psi^{(1)}-3 \Phi^{(1)} \right)\right]=-2\Pi^{(1)}_{ij(em)}\sigma^{ij}_{(1)}-\frac{2}{3} \Delta_{(mag)}^{(1)}\partial_{l}\mathcal{V}_{(1)}^{l} \,,  \nonumber\end{eqnarray}\end{widetext}
where using eqs. (\ref{1fdexpansion}) and  (\ref{2transform})  the energy density magnetic field at second order transforms as
\begin{eqnarray}
\Delta_{(mag)}^{(2)}&=& B^{2}_{(2)} +B^{2\prime}_{(0)} \alpha_{(2)} 
\nonumber\\
&+&  \alpha_{(1)}\left(B^{2\prime \prime}_{(0)}\alpha_{(1)}+B^{2\prime}_{(0)}\alpha_{(1)}^\prime+2B^{2\prime}_{(1)}\right)\nonumber\\
 &+&\xi^i_{(1)}\left(B^{2\prime}_{(0)}\partial_{i}\alpha^{(1)}
+2\partial_{i}B^{2}_{(1)}\right). \label{ultimm}
\end{eqnarray}
The parameters $\alpha$ and $\xi$ are set using the Poisson gauge calculated in appendix \ref{app_evol}. The eq. (\ref{21dynamoequationa}) shows how the field acts as an anisotropic radiative fluid which is important in times where universe is permeated by anisotropic components. In addition, the second term on the right-hand  side describes   the perturbation at first order in the volu\-me expansion.
Equations (\ref{dynamo11}) and (\ref{21dynamoequationa}) show the important role of a magnetized FLRW model.  The set of equations (\ref{dynamo11})-(\ref{2ordendinapoison}) directly offers  a first estimation of how  perturbed four-velocity coupling to magnetic field gives a common  dependence of $B \sim \mathcal{V}^{-\frac{2}{3}}$ under an ideal assumption of infinity conductivity. Hovewer, for a real MHD a complete solution should be calculated together with the case of eq. (\ref{21dynamoequationa}).  The right hand side in eq.(\ref{21dynamoequationa}) provides new phenomenology about the role of the shear and the anisotropic magnetic stress tensor together with a kinematical effect driven for the last term,  reinforcing  the claim in \cite{*20c}. 
In the paper from  Matarrese et al. \cite{*20b} an estimation of the magnetic field to second order dropping the matter anisotropic stress tensor is given by eq.(16) and from this equation they are  to able to compute a solution for the magnetic field, although  in our case we suppose the presence of stress and vector modes at first order possibly generated in early stages from the universe.             
\section{Discussion}\label{conclusi}
A problem in modern cosmology is to explain the origin of cosmic magnetic fields. The origin of these fields is still in debate but  they must affect  the formation of large scale structure and the anisotropies in the cosmic microwave background radiation (CMB) \cite{Tina14,Tina15,Tina16}.  We can see this effect in  eq. (\ref{energy2}) where the evolution of $\Delta_{\mu}^{(2)}$ depends on the magnetic field. In this paper we show that the perturbed metric plays an important role in the global evolution of magnetic fields. 
From our analysis, we wrote a dynamo  like equation for cosmic magnetic fields to second order in perturbation theory in a gauge invariant form. We get the dynamo equation from two approaches. First, using the FLRW as a background space-time and the magnetic fields as a perturbation. The results are eqs. (\ref{dynamo1}) and (\ref{2dynamoequation}) to second order. The second approach a weakly magnetic field was introduced in the background space-time and due to it's statistical properties which allow us to write down the evolution of magnetic field eqs. (\ref{dynamo11}) and (\ref{21dynamoequationa}) and fluid variables in accordance with \cite{tsagas1}. We observe that essentially, the functional form is the same in the two approaches, the coupling between geometrical perturbations and fields variables appear as sources in the magnetic field evolution giving a new possibility to explain the amplification of primordial cosmic magnetic fields.
One important distinction between both approximations is the fractional order in the fields which appears when we consider the magnetic variables as perturbations on the background at difference  when the fields are from the beginning of the background  (section \ref{2part}). Although the first alternative is often used in studies of GWs production in the early universe \cite{o1/2,1/2}, the physical explanation of these fractional orders is sometimes confused, while if we consider an universe permeated with a magnetic density from the background, the  perturbative analysis is more straightforward.
Further studies as anisotropic (Bianchi I) and inhomogeneous (LTB) models
should be addressed to see the implications from the metric behavior in the
evolution of the magnetic field and relax the assumption in the weakness of the
field.  
       
\section{Acknowledgments}
We acknowledge helpful comments from T. Kahniashvili, M. Giovannini, R. Maartens, R. Durrer, and specially to  C. Tsagas for  discussions of the last version. This work was supported by the Observatorio Astron\'omico Nacional from Universidad Nacional de Colombia.
\begin{appendix}
\section{}
\label{app_evol}
For removing the  degrees of freedom we fix the gauge conditions as
 \begin{equation}
 \partial^{i}\omega_{i}^{(r)}=\partial^{i}\chi_{ij}^{(r)}=0,
\end{equation}
this lead to some functions being dropped
 \begin{equation}
 \omega^{\parallel(r)}=\chi^{(r)\bot}_{i}=\chi^{(r)\parallel}=0,
\end{equation}
with the functions defined in eqs. (\ref{omega1}) and (\ref{chi1}). The perturbed metric in the Poisson gauge is given by \ref{metricpoison} thus, using the last constraints  together with eqs. (5.18)-(5.21) in \cite{mollerach} and following the procedure made in \cite{22a}, the vector that  determines the gauge transformation at first order $\xi_{i}^{(1)}=\left(\alpha^{(1)},\partial_{i}\beta^{(1)}+d^{(1)}_{i}\right)$ is given by,
\begin{equation}
\alpha^{(1)}  \rightarrow  \omega_{(1)}^{\parallel}+\beta^{\prime}_{(1)},\quad \beta^{(1)}  \rightarrow -\frac{\chi^{\parallel(1)}}{2},\quad 
d_{i}^{(1)}  \rightarrow -\chi^{\bot(1)}_{i}. \label{apen1}
\end{equation}
Now to second order, when we  use eq. (5.37) in \cite{mollerach} with eq. (\ref{chi1})
we obtain the following transformations
\begin{equation}
\tilde{\chi}^{(2)\parallel} = \chi^{(2)\parallel}+2\beta^{(2)}+\frac{3}{2}\nabla^{-2}\nabla^{-2}\mathrm{X}^{(2)\parallel},
\end{equation}
with
\begin{eqnarray}
&&\mathrm{X}^{(2)\parallel}=2\left(\partial^{i}\partial^{j}D_{ij}\chi_{(1)\parallel}^{\prime}+2H\partial^{i}\partial^{j}D_{ij}\chi^{(1)\parallel}\right)\alpha^{(1)} 
\nonumber \\
&+& \frac{2}{a^{2}}\left(a^{2}\chi_{ij}^{(1)}\right)^{\prime}\partial^{i}\partial^{j}\alpha^{(1)}+2\xi^{k}_{(1)}\partial^{i}\partial^{j}\partial_{k}D_{ij}\chi^{(1)\parallel}
\nonumber \\
&+& 2\partial_{k}\chi^{(1)}_{ij}\partial^{i}\partial^{j}\xi_{(1)}^{k}+2 \left( -4\partial^{i}\partial^{j}\phi_{(1)}+\partial^{i}\partial^{j}\alpha^{(1)}\partial_{0} \right. 
\nonumber\\
&+& \left. \partial^{i}\partial^{j}\xi^{k}_{(1)}\partial_{k} +4H\partial^{i}\partial^{j}\alpha^{(1)}\right)\left(\partial_{(j}d^{(1)}_{i)}+D_{ij}\beta^{(1)} \right)
\nonumber \\
&+& 2\left(-4\phi^{(1)}+\alpha^{(1)}\partial_{0}+\xi^{k}_{(1)}\partial_{k}+4H\alpha^{(1)}\right) \left(\partial^{i}\partial^{j}D_{ij}\beta_{(1)}  \right)    
\nonumber \\
&+& 2\left[ 2\omega_{i}^{(1)}\partial^{i}\nabla^{2}\alpha_{(1)}+2\partial^{j}\nabla^{2}\omega^{\parallel}_{(1)}\partial_{j}\alpha^{(1)}-\partial_{j}\alpha^{(1)}\partial^{j}\nabla^{2}\alpha_{(1)} \right. 
\nonumber \\
&+&\partial^{j}\nabla^{2}\beta^{\prime}_{(1)}\partial_{j}\alpha^{(1)}+\xi_{i}^{(1)\prime}\partial^{i}\nabla^{2}\alpha^{(1)}-\nabla^{2}\left[\left(2\omega^{k}_{(1)}-\partial^{k}\alpha_{(1)} \right. \right. \nonumber \\
&+&\left. \left. \left. \xi^{\prime}_{(1)}\right)\partial_{k}\frac{\alpha_{(1)}}{3}\right]  \right] +    2\left[ 2\partial^{i}\partial^{j}\left( D_{ij}\chi^{\parallel}_{(1)}+\partial_{i}\chi_{k(1)}^{\bot}\right)\partial_{j}\xi^{k}_{(1)} \right. \nonumber \\
&+&  2\chi_{ik}^{(1)}\partial^{i}\nabla^{2}\xi^{k}_{(1)}+2\partial_{j}\xi^{k}_{(1)}\partial^{j}\nabla^{2}\xi_{k}^{(1)} +\partial_{k}\xi_{i}^{(1)}\partial^{i}\nabla^{2}\xi^{k}_{(1)} \nonumber \\
&+& \left. \partial_{j}\xi^{k}_{(1)}\partial^{j}\nabla^{2}\beta_{(1)}-\frac{1}{3}\nabla^{2}\left[ \left( 2\chi_{kl}^{(1)}+2\partial_{(l}\xi_{k)}^{(1)}\right)\partial^{l}\xi^{k}_{(1)}\right] \right].  
\end{eqnarray}
Now if we fix the poisson gauge, $\tilde{\chi}^{(2)\parallel}=0$ we can fix the scalar part of the space gauge
\begin{equation}
 \beta^{(2)}=-\frac{\chi^{(2)\parallel}}{2}-\frac{3}{4}\nabla^{-2}\nabla^{-2}\mathrm{X}^{(2)\parallel}. \label{apen2}
\end{equation}
For the vector space part we should know the transformation rule for the vector part
\begin{equation}
 \tilde{\chi}^{(2)\bot}_{i} = -\partial_{i}\left(\nabla^{-2}\nabla^{-2}\mathrm{X}^{(2)\parallel}\right)+\chi^{(2)\bot}_{i}+d_{i}^{(2)}+                 \nabla^{-2}\mathrm{X}^{(2)\bot}_{i},
\end{equation}
with 
\begin{eqnarray}
&&\mathrm{X}^{(2)\bot}_{i}=2\left( \partial^{j}D_{ij}\chi^{\prime\parallel}_{(1)}+\nabla^{2}\chi^{\prime\bot}_{i(1)}+2H\left(\partial^{j}D_{ij}\chi^{\parallel}_{(1)}  \right. \right. \nonumber \\
&+& \left. \left. \nabla^{2}\chi^{\bot}_{i(1)}\right)\right)\alpha^{(1)}+ \frac{2}{a^{2}}\left(a^{2}\chi_{ij}^{(1)}\right)^{\prime}\partial^{j}\alpha_{(1)}  \nonumber \\
&+&2\xi^{k}_{(1)}\partial_{k}\left(\partial^{j}D_{ij}\chi^{\parallel}_{(1)} +\nabla^{2}\chi_{i}^{\bot(1)}\right)+ 2\partial_{k}\chi_{ij}^{(1)}\partial^{j}\xi^{k}_{(1)}  \nonumber \\
&+& 2\left(-4\partial^{j}\phi_{(1)}+\partial^{j}\alpha_{(1)}\partial_{0}+\partial^{j}\xi^{k}_{(1)}\partial_{k}+ 4H\partial^{j}\alpha^{(1)}\right)\dot  \nonumber \\           
&\dot&\left(\partial_{(j}d^{(1)}_{i)}+D_{ij}\beta^{(1)}\right) +\left(\alpha^{(1)}\partial_{0}-4\phi^{(1)}+\xi^{k}_{(1)}\partial_{k}+4H\alpha^{(1)}\right)\dot \nonumber\ \\
&\dot&\left(\nabla^{2}d_{i}^{(1)}+2\partial^{j}D_{ij}\beta^{(1)}\right)+2\left(\partial^{j}\omega_{i}^{(1)}\partial_{j}\alpha^{(1)}+\omega_{i}^{(1)}\nabla^{2}\alpha^{(1)} \right. \nonumber \\
&+&\nabla^{2}\omega^{\parallel}_{(1)}\partial_{i}\alpha^{(1)}+\omega_{j}^{(1)}\partial^{j}\partial_{i}\alpha^{(1)}-\partial^{j}\partial_{i}\alpha^{(1)}\partial_{j}\alpha^{(1)}  \nonumber \\
&-&\nabla^{2}\alpha^{(1)}\partial_{i}\alpha^{(1)}+\partial^{j}\xi_{i(1)}^{\prime}\partial_{j}\frac{\alpha^{(1)}}{2}+\xi_{i(1)}^{\prime}\nabla^{2}\frac{\alpha^{(1)}}{2} \nonumber \\
&+& \left. \frac{1}{2}\xi_{j(1)}^{\prime}\partial_{i}\partial^{j}\alpha^{(1)}-\frac{1}{3}\partial_{i}\left[\left(2\omega^{k}_{(1)}-\partial^{k}\alpha_{(1)}+\xi^{\prime}_{(1)}\right)\partial_{k}\alpha_{(1)}\right]\right) \nonumber \\
&+& 2\left(\chi_{ik}^{(1)}\nabla^{2}\xi^{k}_{(1)}+\partial_{i}\xi^{k}_{(1)}\left( \partial^{j}D_{jk}\chi^{\parallel}_{(1)}+\nabla^{2}\chi_{k}^{(1)\bot}\right) \right. \nonumber \\  
&+&\chi_{jk}^{(1)}\partial^{j}\partial_{i}\xi^{k}_{(1)}+\partial_{j}\xi^{k}_{(1)}\partial^{j}\partial_{i}\xi_{k}^{(1)}+\nabla^{2}\xi^{k}_{(1)}\partial_{i}\xi_{k}^{(1)} \nonumber \\
&+&\frac{1}{2}\partial^{j}\partial_{k}\xi_{i}^{(1)}\partial_{j}\xi^{k}_{(1)} +\frac{1}{2}\partial_{k}\nabla^{2}\beta^{(1)}\partial_{i}\xi^{k}_{(1)}+\frac{1}{2}\partial_{k}\xi_{j}^{(1)}\partial^{j}\partial_{i}\xi^{k}_{(1)} \nonumber \\
&+&\nabla^{2}\beta^{\prime}_{(1)}\partial_{i}\frac{\alpha^{(1)}}{2}+\frac{1}{2}\partial_{k}\xi_{i}^{(1)}\nabla^{2}\xi^{k}_{(1)}\nonumber \\ 
&+& \partial^{j}\chi_{ik}^{(1)}\partial_{j}\xi^{k}_{(1)}-\left. \frac{1}{3}\partial_{i}\left[ \left( 2\chi_{kl}^{(1)}+2\partial_{(l}\xi_{k)}^{(1)}\right)\partial^{l}\xi^{k}_{(1)}\right]\right).     
\end{eqnarray}
Now we use the condition $\tilde{\chi}^{(2)\bot}_{i}=0$ for instance,
\begin{equation}
d_{i}^{(2)} = \partial_{i}\left(\nabla^{-2}\nabla^{-2}\mathrm{X}^{(2)\parallel}\right)-\chi^{(2)\bot}_{i}-                 \nabla^{-2}\mathrm{X}^{(2)\bot}_{i}. \label{apen3}
\end{equation}
To find the temporal part of the gauge transformation, we use the eq. (5.35) in \cite{mollerach} and eq. (\ref{omega1}).
With some algebra,  the scalar part  transforms like
\begin{equation}
 \tilde{\omega}^{(2)\parallel}=\omega^{(2)\parallel}-\alpha^{(2)}+\beta^{\prime}_{(2)}+\nabla^{-2}\mathrm{W}^{(2)\parallel}, 
\end{equation}
with
\begin{eqnarray}
&& \mathrm{W}^{(2)\parallel} =   -4\left(\partial^{i}\psi^{(1)}\partial_{i}\alpha^{(1)}+\psi^{(1)}\nabla^{2}\alpha^{(1)}\right)+\partial^{i}\alpha^{(1)}\left[ 2\omega_{i}^{(1)\prime} \right. \nonumber \\
&+& \left. 4H\omega_{i}^{(1)}-\partial_{i}\alpha^{\prime}_{(1)}+\xi^{\prime\prime}_{(1)i}-4H\left(\partial_{i}\alpha^{(1)}-\xi_{i(1)}^{\prime}\right)\right] \nonumber \\
&+& \alpha^{(1)}\left[2\nabla^{2}\omega^{\prime}_{(1)\parallel} +4H\nabla^{2}\omega^{(1)}_{\parallel}-\nabla^{2}\alpha^{\prime}_{(1)}+\nabla^{2}\beta^{\prime\prime} \right. \nonumber \\
&-& \left. 4H\left(\nabla^{2}\alpha^{(1)}-\nabla^{2}\beta^{\prime}_{(1)}\right)\right]+\partial^{i}\xi^{j}_{(1)}\left(2\partial_{j}\omega_{i}^{(1)} \right. \nonumber \\
&-& \left. \partial_{i}\partial_{j}\alpha^{(1)}+\partial_{j}\xi^{\prime}_{i(1)}\right)+\xi^{j}_{(1)}\left(2\partial_{j}\nabla^{2}\omega^{\parallel}_{(1)}-\partial_{j}\nabla^{2}\alpha^{(1)}  \right.  \nonumber \\             
&+& \left.  \partial_{j}\nabla^{2}\beta^{\prime}_{(1)}\right)+\alpha^{\prime}_{(1)}\left(2\nabla^{2}\omega^{\parallel}_{(1)}-3\nabla^{2}\alpha^{(1)}+\nabla^{2}\beta^{\prime}_{(1)}\right) \nonumber \\   
&+& \partial^{i}\alpha^{\prime}_{(1)}\left(2\omega_{i}^{(1)}-3\partial_{i}\alpha^{(1)}+\xi^{\prime}_{(1)i}\right)+\nabla^{2}\xi^{j}_{(1)}\left(2\omega_{j}^{(1)} \right. \nonumber\\
&-& \left. \partial_{j}\alpha^{(1)}\right)+\partial_{i}\xi^{j}_{(1)}\left(2\partial^{i}\omega_{j}^{(1)}-\partial^{i}\partial_{j}\alpha^{(1)}\right)+\partial^{i}\xi^{j\prime}_{(1)}\left[\partial_{j}\xi_{i}^{(1)} \right.  \nonumber\\
&+& \left.  2\chi_{ij}^{(1)}+2\partial_{i}\xi_{j}^{(1)}-4\phi^{(1)}\delta_{ij}\right]+\xi^{j\prime}_{(1)}\left[-4\partial_{j}\phi^{(1)}  \right. \nonumber \\
&+&\left. 2\left(\partial^{i}D_{ij}\chi^{\parallel}_{(1)} + \nabla^{2}\chi_{j}^{(1)\bot}+2\nabla^{2}\xi_{j}^{(1)}+\partial^{i}\partial_{j}\xi_{i}^{(1)}\right) \right],    
\end{eqnarray}
in this way we  fix the temporal part of the gauge using $\tilde{\omega}^{(2)\parallel}=0$ in the last equation finding the follows 
\begin{equation}
 \alpha^{(2)}=\omega_{(2)}^{\parallel}+\partial_{0}\beta_{(2)}+\nabla^{-2}\mathrm{W}^{(2)\parallel}. \label{apen4}
\end{equation}
Therefore, we found explicitly the set of functions that  fix the gauge dependence given by eqs. (\ref{apen1}),(\ref{apen2}),(\ref{apen3}) and (\ref{apen4}).  Thus, using the above equations we can calculate the gauge dependence in the scalar perturbations at second order that  were shown in eq. (\ref{psi2ordergi})
\begin{equation}
\mathcal{S}_{(2)}^{||}=\omega_{(2)}^{\parallel}-\frac{\chi^{\parallel(2)\prime}}{2}-\frac{3}{4}\nabla^{-2}\nabla^{-2}\mathrm{X}^{(2)\prime}_{\parallel}+\nabla^{-2}\mathrm{W}^{(2)\parallel},
\end{equation}
which can be interpreted like shear to second order, again we see the last equation is a generalization for the first order scalar shear plus quadratic terms in the perturbed functions.  
\section{}
\label{app_evol2}
To find the charge evolution, we use the fact that $j^{\alpha}_{;\alpha}=0$ therefore, the temporal part of this equation drive us to the charge conservation
\begin{equation}
 \varrho^{(1)\prime}+3H\varrho^{(1)}+\partial_{i}J^{i}_{(1)}=0,
\end{equation}
at first order in the approximation and
 \begin{eqnarray}
\varrho^{(2)\prime}&+&3H\varrho^{(2)}+\partial_{i}J^{i}_{(2)}+\left(\Psi^{(1)\prime}-3\Phi^{(1)\prime}\right)\varrho^{(1)} \nonumber \\ &+&\partial_{i}\left(\Psi^{(1)\prime}-3\Phi^{(1)\prime}\right)J^{i}_{(1)}=0 ,
\end{eqnarray}
to second order. These equations are important for resol\-ving the dynamo equation.  
In the section \ref{secapedb} was found the  momentum equation at second order, where $\boldsymbol{S}^{13}_{i}$ is given by 
\begin{eqnarray}
\boldsymbol{S}^{13}_{i}&=&\frac{\left[\Delta^{(1)}\left(1+c_{s}^{2}\right)\mathcal{V}_{i}^{(1)}\right]^{\prime}}{\mu_{(0)}\left(1+w\right)}+4H\frac{\Delta^{(1)}\left(1+c_{s}^{2}\right)\mathcal{V}_{i}^{(1)}}{\mu_{(0)}\left(1+w\right)}\nonumber \\
&+&2\frac{\left[\mu_{(0)}\left(1+w\right)\chi_{ij}^{(1)}\upsilon_{(1)}^{j}\right]^{\prime}}{\mu_{(0)}\left(1+w\right)}+8H\chi_{ij}^{(1)}\upsilon_{(1)}^{j}-8H\Phi^{(1)}\upsilon_{i}^{(1)} \nonumber \\
&-&\frac{\left[\mu_{(0)}\left(1+w\right)\left(\omega_{i}^{(1)}+\mathcal{V}_{i}^{(1)}\right)\right]^{\prime}\Psi^{(1)}}{\mu_{(0)}\left(1+w\right)}+\frac{\Delta^{(1)}\left(1+c_{s}^{2}\right)\partial_{i}\Psi^{(1)}}{\mu_{(0)}\left(1+w\right)}\nonumber \\
&-&\omega_{i}^{(1)}\Psi^{\prime}_{(1)}-4H\Psi^{(1)}\left(\mathcal{V}_{i}^{(2)}+\omega_{i}^{(1)}\right) +\upsilon^{j}\left(\partial_{j}\upsilon_{i}^{(1)} \right)\nonumber \\
&-&3\Phi^{\prime}_{(1)}\left( \mathcal{V}_{i}^{(1)}+\upsilon_{i}^{(1)} \right)-\upsilon_{(1)}^{j}\partial_{[i}\omega_{j]}^{(1)}-2\Psi^{(1)}\partial_{i}\Psi^{(1)} \nonumber \\
&-&2\Phi^{(1)}\left[\mu_{(0)}\left(1+w\right)\upsilon^{(1)}_{i}\right]^{\prime}+\frac{\left(\partial_{j}\Psi^{(1)}+H\omega_{j}^{(1)}\right) \Pi^{j(1)}_{i}}{\mu_{(0)}\left(1+w\right)} \nonumber\\
&+&\mathcal{V}_{i}^{(1)}\partial_{\alpha}\partial^{\alpha}\upsilon_{(1)}-\frac{6\partial_{j}\Phi^{(1)}\Pi^{j(1)}_{i}-\frac{1}{2}\partial_{j}\chi^{k(1)}_{i}\Pi^{j(1)}_{k} }{\mu_{(0)}\left(1+w\right)},
\end{eqnarray}
where $w=\frac{P_{(0)}}{\mu_{(0)}}$ is the state equation  ($w=0$ for dust  and  $w=1/3$ for radiation era)  and $c_{s}^{2}=\frac{P_{(1)}}{\mu_{(1)}}$ the adiabatic sound speed. 
Using the expression for the momentum exchange among particles and the momentum conservation, we obtain the following equations for protons, electrons and photons during radiation era  
\begin{eqnarray}
&&\mu_{(0)}^{(p)}\left[\mathcal{V}_{i}^{(2)(p)\prime}+H\mathcal{V}_{i}^{(2)(p)}+\partial_{i}\Psi^{(2)}+2\boldsymbol{S}^{13(p)}_{i}\right]+\partial_{i}\Delta_{P}^{(2)(p)}\nonumber \\
&+&\frac{4}{3}\partial_{i}\nabla^{2}\Pi^{(2)}_{(p)}=a^{4}\left(E_{i}^{(1)}\varrho_{(1)}^{(p)}+\epsilon_{ij}^{k}J_{(1)}^{j(p)}B_{k}^{(1)} \right)+K^{(ep)}_{i},\label{proton}
\end{eqnarray}
\begin{eqnarray}
&&\mu_{(0)}^{(e)}\left[\mathcal{V}_{i}^{(2)(e)\prime}+H\mathcal{V}_{i}^{(2)(e)}+\partial_{i}\Psi^{(2)}+2\boldsymbol{S}^{13(e)}_{i}\right]\nonumber \\
&+&\partial_{i}\Delta_{P}^{(2)(e)}+\frac{4}{3}\partial_{i}\nabla^{2}\Pi^{(2)}_{(e)} \nonumber \\
&=&a^{4}\left(E_{i}^{(1)}\varrho_{(1)}^{(e)}+\epsilon_{ij}^{k}J_{(1)}^{j(e)}B_{k}^{(1)} \right)-K^{(ep)}_{i}+K^{(\gamma)}_{i},
 \label{electron}\end{eqnarray}      
\begin{eqnarray}
&&\frac{4}{3}\mu_{(0)}^{(\gamma)}\left[\mathcal{V}_{i}^{(2)(\gamma)\prime}+\partial_{i}\Psi^{(2)}+2\boldsymbol{S}^{13(\gamma)}_{i}\right]+\partial_{i}\Delta_{P}^{(2)(\gamma)}\nonumber \\
&+&\frac{4}{3}\partial_{i}\nabla^{2}\Pi^{(2)}_{(\gamma)}=-K^{(\gamma)}_{i}, \label{photon}\end{eqnarray}

\end{appendix}
\nocite{*}
\bibliography{Main_text_file}
\end{document}